\RequirePackage{luatex85}
\documentclass[twocolumn,amsmath,amssymb,aps,pra,superscriptaddress,showpacs,10pt]{revtex4-2}

\usepackage{graphicx}
\usepackage{grffile} 
\usepackage{array}
\usepackage{bbm}
\usepackage{color}
\usepackage[utf8]{inputenc}
\usepackage[OT1]{fontenc}
\usepackage{shellesc}
\usepackage{stringenc}

\usepackage{units} 

\usepackage{upgreek} 

\usepackage{newtxtext} 
\usepackage{newtxmath}

\usepackage{microtype}

\usepackage{dsfont}
\usepackage{xcolor}

\newcommand{\W}{\text{W}}

 \newcommand{\bra}[1]{\left\langle{#1}\right|}
 \newcommand{\ket}[1]{\left|{#1}\right\rangle}
 \newcommand{\braket}[2]{\langle{#1}|{#2}\rangle}

\newcommand{\e}[1]{\operatorname{e}^{#1}}

\newcommand{\I}{i}

\newcommand{\D}{\text{d}}

\def\ba#1\ea{\begin{align}#1\end{align}}																

\usepackage{braket}

\usepackage{tikz}                       
\usepackage{pgfplots}                   
\usetikzlibrary{spy}
\usetikzlibrary{external}
\tikzset{external/system call={lualatex -shell-escape -file-line-error -halt-on-error 
-interaction=batchmode -jobname "\image" "\texsource"}}
\tikzsetexternalprefix{Bilder/}
\usetikzlibrary{shapes,arrows,shadows,3d,calc}
\usetikzlibrary{decorations.pathmorphing,decorations.pathreplacing,decorations.markings,trees}
\usetikzlibrary{pgfplots.groupplots}

\pgfplotsset{%
  every axis legend/.append style={%
    cells={anchor=west},
    at={(0.96,0.04)},
    anchor=south east,
    font=\scriptsize
  },
  every axis/.append style={%
    yticklabel style={/pgf/number format/fixed zerofill, /pgf/number format/precision=2}
  },
  width= 0.45\textwidth, height=5cm, xmajorgrids=false, xminorgrids=false, minor x tick num=1
}

\usepackage[hang]{subfigure} 

\pgfplotsset{compat=1.8}

\usepackage{blindtext} 
\usepackage{booktabs}

\newcommand*{\balancecolsandclearpage}{%
  \close@column@grid
  \cleardoublepage
  \twocolumngrid
}

\definecolor{cset-aps-blueberry}{RGB}{28,128,158}
\definecolor{cset-aps-blue}{RGB}{46,44,184}
\definecolor{cset-aps-turquoise}{RGB}{0,67,88}
\definecolor{cset-aps-limegreen}{RGB}{190,219,67}
\definecolor{cset-aps-green}{RGB}{31,138,112}
\definecolor{cset-aps-yellow}{RGB}{255,225,25}
\definecolor{cset-aps-orange}{RGB}{253,116,0}
\definecolor{cset-aps-red}{RGB}{219,0,43}
\usepackage{hyperref}
\hypersetup{%
	colorlinks=true,
	linkcolor={cset-aps-red},
	linkbordercolor={cset-aps-red},
	filecolor={cset-aps-orange},
	filebordercolor={cset-aps-orange},
	citecolor={cset-aps-blue},
	citebordercolor={cset-aps-blue},
	urlcolor={cset-aps-green},
	urlbordercolor={cset-aps-green},
	menucolor={cset-aps-limegreen},
	menubordercolor={cset-aps-limegreen},
	breaklinks=true,
	pdfborderstyle={/S/U/W 2},
	pdfpagemode=UseOutlines,
	pdfstartpage={1},
}

\begin{document}
\title{Multiphoton processes and higher resonances in the quantum regime of the free-electron laser  }
  \author{Peter Kling}
  \affiliation{German Aerospace Center (DLR), Institute of Quantum Technologies, Wilhelm-Runge-Stra{\ss}e 10, D-89081 Ulm, Germany}
\author{Enno Giese}
\address{Technische Universit{\"a}t Darmstadt, Fachbereich Physik, Institut f{\"u}r Angewandte Physik, Schlossgartenstr. 7, D-64289 Darmstadt, Germany}

\begin{abstract}
Despite exhibiting novel radiation features, the operation of the proposed quantum free-electron laser would have the drawback that the number of emitted photons is limited by one per electron, significantly reducing the output power of such a device. We show that relying on different resonances of  the initial momentum of the electrons increases the number of emitted photons, but also increases the required length of the undulator impeding an experimetal realization. Moreover, we investigate how multiphoton processes influence the dynamics in the deep quantum regime.                
\end{abstract}

\maketitle

\section{Introduction}
\label{sec:Introduction}

The quantum free-electron laser (Quantum FEL)~\cite{schroeder,pio_prl,boni17,serbeto09,NJP2015,brown17,schaap2022} is a proposed radiation source which shows outstanding radiation features in the x-ray regime~\cite{boni06,PRR2021} and is anticipated to be a useful tool for applications in material and life sciences~\cite{debus,Seddon2017}.
We focused in recent studies~\cite{NJP2015,PRA2019,PRR2021} on single-photon scattering to describe the dynamics of the system.
In this paper we complement these studies and show how multiphoton processes as well as different resonances of the initial electron momentum affect the FEL dynamics and we discuss their consequences for an experimental realization.

According to Ref.~\cite{boni_epl} the occurrence of higher-order resonances and the resulting dynamics would be absent in a semiclassical model. 
In contrast, we offer an elementary explanation for higher resonances in terms of energy-momentum conservation that is still captured by the semi-classical Hamiltonian. 

The underlying mechanism of FEL physics is Compton scattering~\cite{bosco}, where an electron absorbs a wiggler photon and emits a laser photon -- or the vice versa process. Consequently, the momentum $p$ of the  electron changes by a discrete recoil $q\equiv 2\hbar k$, where $\hbar$ represents the reduced Planck constant and $k=k_{\text{L}}=k_{\W} $ is the wave number of the laser and the wiggler field in the co-moving Bambini--Renieri frame~\cite{bambi,*brs}.

\begin{figure}
   \centering 
      \includegraphics[scale=1]{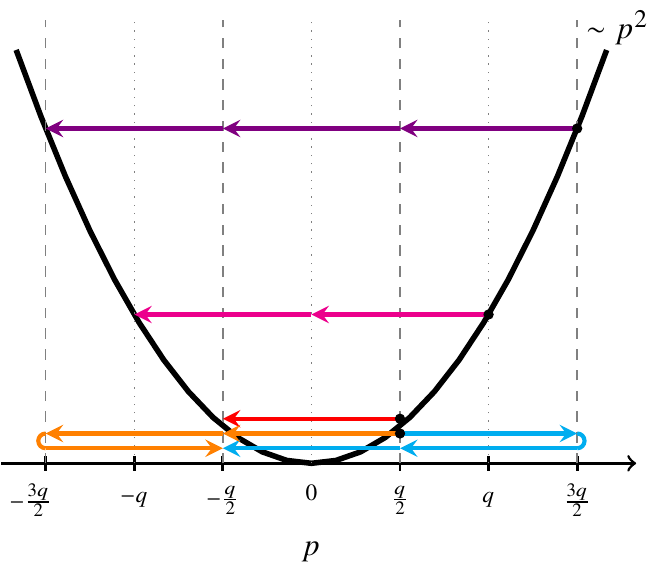}
  \caption{
Resonant transitions in an FEL visualized by energy-momentum conservation: We have drawn the kinetic energy $\sim p^2$ (parabola) of an electron as a function of the momentum $p$ in the Bambini--Renieri frame~\cite{bambi,*brs}, where the wave numbers of the laser and the wiggler mode coincide, that is $k\equiv k_{\text{L}}=k_{\W}$ and the motion of the electron is non-relativistic. In an elastic Compton-scattering event a wiggler photon is absorbed and a laser photon is emitted, or vice versa. Hence, (i) the momentum of the electron changes by multiples of the recoil $q\equiv 2\hbar k$ and (ii) the total kinetic energy of  electron and photons has to be conserved. The first condition implies that the distance between initial and finite momenta has to be an integer multiple of $q$. The second condition means that only transitions are allowed that horizontally connect two points on the energy parabola. (These points have the same distance from the $x$-axis due to our specific frame of reference.) These two conditions are only fulfilled by initial and final momenta of the form $p=\nu q/2$ with $\nu$ being an integer.  We consider single-, two-, and three-photon transitions from the three lowest resonant momenta, that is $p=q/2$, $p=q$, and $p=3q/2$ to a different resonant momentum. 
  For the first resonance, the transition from $q/2$ to $-q/2$ is resonant, which can be achieved by the emission of a single photon or via three-photon processes, where two photons are emitted and one photon is absorbed. Regarding the second resonance there are no resonant transitions with an odd number of photons. However, transitions with an even number of photons can be resonant, for example two-photon processes between $q$  and $-q$. For the third resonance, we require at least three momentum steps to connect the momenta $3q/2$ and $-3q/2$. We note that the situation is mirrored for the momenta $-q/2$, $-q$, and $-3q/2$, with photon emission interchanged with absorption.} 
 \label{fig:energy_momentum}
 \end{figure}

During such an elastic scattering event not only the total momentum  has to be conserved, but also the kinetic energy $\sim p^2$.
From energy-momentum conservation we obtain (also higher-order) resonances for the initial momentum at integer multiples of $q/2$. The emergence of these resonances is visualized in Fig.~\ref{fig:energy_momentum} by identifying  the resonant transitions with the help of energy parabolas in momentum space:         

The first resonant process at $p=q/2$ occurs when the electron resonantly emits \textit{one} laser photon 
and it jumps to  the momentum $-q/2$. By the inverse process the electron can return to $q/2$ resulting in a two-level system, which we identified in Ref.~\cite{NJP2015} as Quantum FEL in accordance with Ref.~\cite{boni06}.
In contrast, for $p=q$ there is no resonant single-photon transition.
However, the electron can take two steps on the momentum ladder from $q$ to $-q$ while emitting \textit{two} laser photons.             

At first sight, such higher resonances seem favorable since more emitted photons imply a higher output intensity.
However, the typical timescale of the dynamics increases for higher resonances~\cite{boni_epl,schaap2022}.
A longer interaction time requires a longer undulator and thus adds additional challenges to an experimental realization of a Quantum FEL~\cite{steiniger}.
Moreover, damping mechanisms like spontaneous emission~\cite{robb2012} or space-charge effects~\cite{loui78,sprangle1,schaap2022} destroy an efficient Quantum FEL operation already for relatively small interaction times~\cite{debus}.

According to Fig.~\ref{fig:energy_momentum} the number of involved momentum steps and by that the number of emitted/absorbed photons increases for higher-order resonances. 
Probabilities for multiphoton processes scale in general with powers of the coupling strength between light and matter.
Specifically, in the quantum theory of the FEL this behavior implies a scaling in powers of the quantum parameter, that is the ratio of the coupling strength  to the recoil.
For quantum effects to emerge, this parameter has to be small~\cite{NJP2015} and thus multiphoton transitions are suppressed when compared to the single-photon processes at $p=q/2$. 
In this paper we prove this behavior  by employing the method of averaging over rapid oscillations~\cite{bogoliubov,higher} in the low-gain regime (Sec.~\ref{sec:Low-gain_FEL}) as well as in the high-gain regime (Sec.~\ref{sec:High-gain_FEL}) of FEL operation.

In App.~\ref{sec:Effective_Hamiltonian} we derive the effective Hamiltonian of our asymptotic method. While we deal in App.~\ref{sec:Population_of_Momentum_Levels} with the population of the momentum levels in the low-gain regime, we show in App.~\ref{sec:Calculations_in_High-Gain_Regime} our calculations in the high-gain regime.

\section{Low-gain FEL}
 \label{sec:Low-gain_FEL}

In the low-gain regime of FEL operation~\cite{schmueser} the mean photon number $n$ changes only marginally during the interaction with an electron bunch and the motion of an electron decouples from the motion of the others~\cite{peter}.
Hence, we restrict ourselves to the quantized motion of a single electron with mass $m$ coupled to a classical and fixed radiation field.

The motion of an electron with initial momentum $p$ may only change by  integer multiples of the recoil $q$. We describe the resulting momentum ladder through the momentum jump operator 
      \begin{equation}
           \hat{\sigma}_{\mu,\nu}\equiv \ket{p-\mu q}\bra{p - \nu q}
       \end{equation}
with $\mu$ and $\nu$ being integers.

In Ref.~\cite{NJP2015} we  defined the quantum parameter $\alpha_n \equiv g\sqrt{n}/\omega_\text{r}$ as the ratio of the coupling strength $g\sqrt{n}$ and the recoil frequency $\omega_\text{r}\equiv q^2/(2m\hbar)$.
For quantum effects to emerge we require (i) that the quantum parameter is small, that is $\alpha_n \ll 1$, and (ii) that the initial moment spread $\Delta p$ of the electron beam is small, that is $\Delta p \ll q$.
Else, the discrete motion of the electron is washed out and the particle follows continuous trajectories~\cite{NJP2015,carmesin20}.
Throughout this paper, we assume for simplicity that the electron is initially described by a momentum eigenstate $\ket{p}$.    
            
The asymptotic method of averaging separates the resonant processes from the non-resonant ones.
For the former ones we formulate an effective Hamiltonian $\hat{H}_\text{eff}$~\cite{higher} and asymptotically expand it in powers of $\alpha_n$.
We  solve the resulting Schr{\"o}dinger equation exactly, which gives rise to slowly-varying part of the dynamics.
For the non-resonant transitions we rely on a perturbative solution which leads to amplitude corrections including rapidly varying terms.
Each additional step on the momentum ladder raises the order of the asymptotic expansion by one. 
 
   \begin{figure}
            \includegraphics[scale=1]{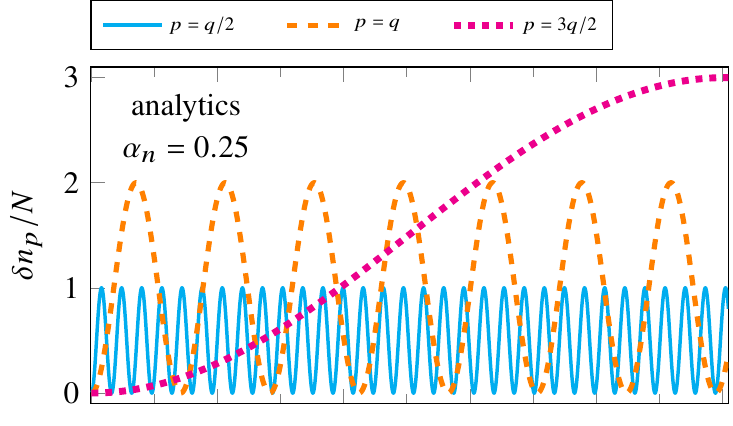}
            \includegraphics[scale=1]{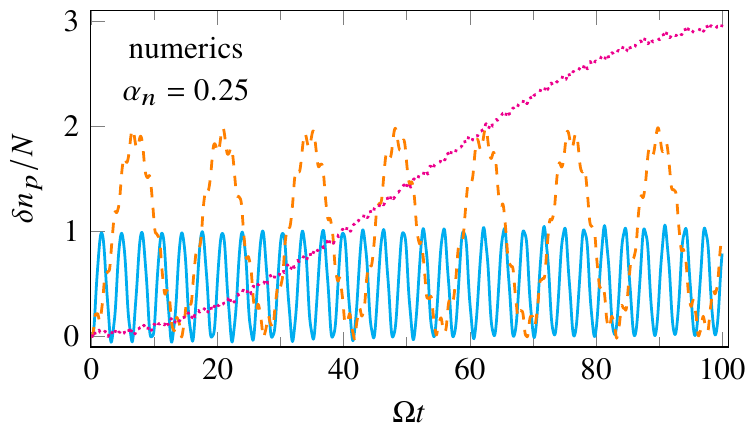}
     \caption{Change $\delta n_p$ of the mean photon number in a low-gain FEL in the quantum regime divided by the number $N$ of electrons as a function of the phase $\Omega t$ with the Rabi frequency $\Omega$ for the first resonance at $p=q/2$. We compare the curves for three different initial electron momenta, that is (i) $p=q/2$ (cyan line), (ii) $p=q$ (orange, dashed line), and (iii) 
$p=3q/2$ (magenta, dotted line) for a fixed value of the quantum parameter of $\alpha_n=0.25$. If we increase the order of the  resonance, the number of  emitted photons per electron increases. However, the photon number grows more slowly for higher resonances. We observe that the analytical results (top) from Eq.~\eqref{eq:dn_low}  and the numerical simulation (bottom) agree.}
     \label{fig:time_scales}
   \end{figure}

In the following, we consider the change of the mean photon  number  $\delta n_p (t)\equiv \braket{\hat{n}(t)}-\braket{\hat{n}(0)}$ during the interaction of an electron bunch containing $N$ electrons of momentum $p$ with the fields.
In the low-gain regime this change has to be smaller than the initial photon number $n_0\equiv\braket{\hat{n}(0)}$, that is $\delta n_p \ll n_0$.
Since each momentum step translates to the emission or absorption of  a photon,  we calculate the change $\delta n_p$  via the relation 
\begin{equation}
\delta n_p (t)=N\sum\limits_\mu \mu \, P_{p-\mu q}(t),
\end{equation}
where $P_{p-\mu q}$ denotes the time-dependent probability that the momentum level $p-\mu q$ is populated.

According to App.~\ref{sec:Population_of_Momentum_Levels} we find that for the initial condition $p= \nu q/2$, the population of levels $\pm \nu q/2$ corresponding to resonant transitions are described by Rabi oscillations between zero and unity, while the probabilities corresponding to non-resonant transitions are suppressed with powers of $\alpha_n$.  
With the help of the explicit expressions for the $P_{p-\mu q}$ in App.~\ref{sec:Population_of_Momentum_Levels} we arrive at the results 
\begin{subequations}\label{eq:dn_low}
\begin{align}
\delta n_{q/2}(t)&\cong N \sin^2{\left[\Omega t \left(1-\frac{\alpha_n^2}{4}\right)\right]}\,,\\
\delta n_q (t)&\cong 2 N \sin^2{\left[\alpha_n\Omega t \left(1-\frac{16\alpha_n^2}{9}\right)\right]}\,, \ \ \   \text{and}
\\
\delta n_{3q/2}(t) &\cong 3N \sin^2{\left(\frac{\alpha_n^2}{4}\Omega t \right)}
\end{align}
\end{subequations}
of $\delta n_p$ for the first, second, and third resonance, where we have defined the Rabi frequency $\Omega\equiv g\sqrt{n}$ of the fundamental resonance $q/2$.
Here we have only included the leading orders in amplitude and the lowest-order corrections in frequency.  
We obtain that for higher resonances (i) the number of maximally emitted photons increases, but also that (ii) the effective Rabi frequency becomes smaller leading to a slower growth of the mean photon number as apparent from Fig.~\ref{fig:time_scales}.  

The calculation of higher-order resonances requires higher orders of the asymptotic expansion and consequently this increase of time scales continues beyond the third resonance. Hence, we  expect  the scaling
  \begin{equation}
   \Omega^{(\nu)} \propto \alpha_n^{\nu-1} \Omega
  \end{equation}
for the effective Rabi frequency $\Omega^{(\nu)}$ that corresponds to the resonant transition from $\nu q/2$ to $-\nu q/2$~\cite{peter}.
We emphasize that the emergence of different time scales for different initial momenta follows directly from the number of  momentum steps necessary for a resonant transition. An analogous behavior has been also observed in atomic diffraction~\cite{Kunze1996,Ahlers2016}.
However, since we observe this dynamics in a semi-classical model, it has nothing to do with a quantized light field in contrast to the assumption of Ref.~\cite{boni_epl}. 

\section{High-gain FEL}
 \label{sec:High-gain_FEL}
 
In the high-gain regime of FEL operation, the relative change of the laser intensity during the interaction with the electrons is large and consequently the laser field cannot be seen as an fixed, external field. In contrast, the motion of each electron in the bunch influences the motion of the remaining electrons via their common interaction with the laser field~\cite{boni_coll,PRA2019}. 

In analogy to Ref.~\cite{PRA2019} we employ a collective model, where the single-particle jump operators are replaced by their collective counterparts, that is        
  \begin{equation}
  \label{eq:collective}
   \hat{\sigma}_{\mu,\nu} 
   \rightarrow  \hat{\Upsilon}_{\mu,\nu}\equiv\sum\limits_{j=1}^N \hat{\sigma}_{\mu,\nu}^{(j)}\,,
  \end{equation}
where $\hat{\sigma}_{\mu,\nu}^{(j)}$ is the single-particle operator for electron $j$.
We assume that each electron is initially described by a momentum eigenstate with the same momentum $p$ yielding the product state $\ket{p,p,...,p}$. Moreover, we introduce a quantized laser mode with the bosonic annihilation and creation operators, respectively, satisfying the commutation relation $\displaystyle\left[\hat{a}_{\text{L}},\hat{a}_{\text{L}}^\dagger\right]=1$. For the calculation of the mean photon number we restrict ourselves to an FEL seeded by a Fock state with $n_0$ photons.    
 
In Refs.~\cite{PRA2019,PRR2021} we found that the leading order of the effective Hamiltonian 
for the first resonance $p=q/2$ is given by the Dicke Hamiltonian, which describes the collective interaction of many two-level atoms with a quantized mode of the radiation field~\cite{dicke}. 

In the current paper, we include  the lowest-order corrections emerging from the higher orders of $\hat{H}_\text{eff}$ derived in App.~\ref{sec:Effective_Hamiltonian}. From the results in Ref.~\cite{PRR2021} and from Eqs.~\eqref{eq:cdot} and~\eqref{eq:aq2} we deduce for $p=q/2$ the approximate expression 
\begin{equation}
\label{eq:nkl_erste}
n_{q/2}(L)\!=\!n_0 + N \text{cn}^2{\left[\!\sqrt{1+\frac{n_0}{N}}\frac{L}{2L_g}\!
\left[1-\frac{\alpha_N^2}{8}\!\left(1+\frac{2n_0}{N}\right)\!\right]
\!-\!K,\mathfrak{K} \right]}
\end{equation}
for the mean photon number $n_{p/2}$ as a function of the undulator length $L\equiv ct$. Here $c$ denotes the velocity of light and $L_g\equiv c/(2g\sqrt{N})$ represents the gain length of a Quantum FEL~\citep{boni06,PRA2019}. The Jacobi elliptic function $\text{cn}$  depends on its modulus        
$\displaystyle \mathfrak{K}\equiv\left(1+n_0/N\right)^{-1/2}$ and $K\equiv K(\mathfrak{K})$ denotes the corresponding complete elliptic integral of first kind~\cite{byrd}. We note that the quantum parameter $\alpha_N \equiv g \sqrt{N}/\omega_\text{r}$ for the high-gain regime depends on the number $N$ of electrons in the bunch.

\begin{figure}
 \centering
  \includegraphics[scale=1]{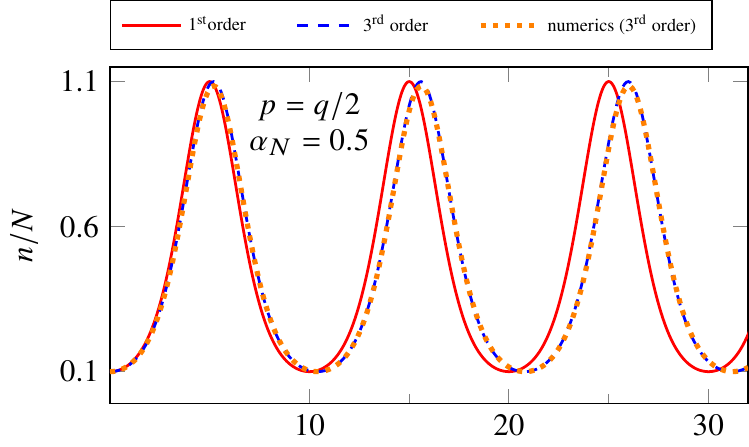}\\
  \includegraphics[scale=1]{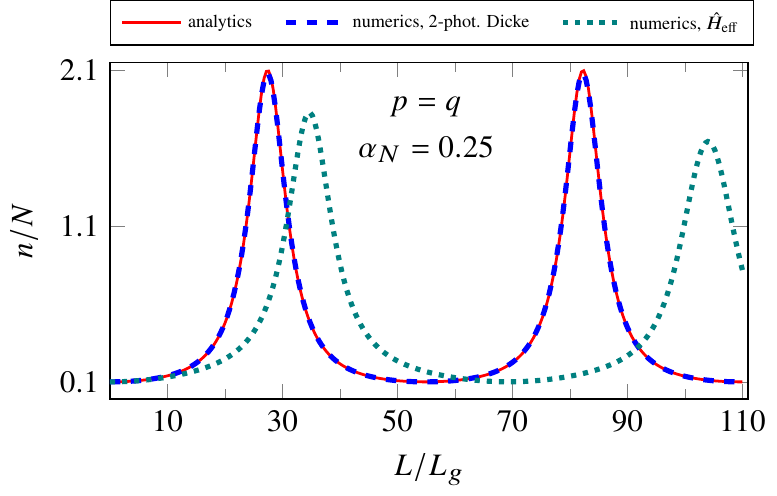}
 \caption{Mean photon number $n$ of a seeded high-gain FEL in the quantum regime divided by the number $N$ of electrons as a function of the undulator length $L$ in units of the gain length $L_g$. The initial photon number amounts to $n_0=0.1 N$ and the electron number to $N=10^4$. In the top panel all electrons start at the first resonance $p=q/2$ and we have chosen the value $\alpha_N=0.5$ for the quantum parameter. We observe that the analytical solution(blue, dashed line)  from Eq.~\eqref{eq:nkl_erste}  including third-order corrections agrees with the numerical solution corresponding to the effective Hamiltonian in third order (orange, dotted line), while the first-order solution (red line) of Ref.~\citep{PRR2021}  differs by a phase shift $\sim \alpha_N^2$. 
In the bottom panel all electrons start at the second resonant momentum $p=q$ with $\alpha_N=0.25$. Here we compare the analytical approximation (red line) from Eq.~\eqref{eq:qhigh_nkl_zweite} to the numerical simulations resulting (i) from the two-photon Dicke Hamiltonian (blue, dashed line), and (ii) from the full effective Hamiltonian (green, dotted line) of second order.
In all three cases we observe an oscillatory behavior, with at most two emitted photons per electron. Analytics  and numerics agree for the simplified model, that is the two-photon Dicke Hamiltonian. However, the simulation for the full dynamics shows a suppressed maximum photon number which occurs after a higher interaction length in comparison to the curves corresponding to the simplified model.
Nevertheless, the qualitative behavior is similar.}
 \label{fig:n_q}
\end{figure}

In the top panel of Fig.~\ref{fig:n_q} we compare the approximation for $n$ to the numerical simulation corresponding to the effective Hamiltonian up to third order. For $\alpha_N \ll 2\sqrt{2}$ the phase corrections in Eq.~\eqref{eq:nkl_erste} are negligible and thus we obtain only a small phase shift for $\alpha_N=0.5$ between third-order and first-order results of the asymptotic method of averaging.
While this frequency shift is perfectly predicted by Eq.~\eqref{eq:nkl_erste}, numerics reveals a very small suppression of the amplitude which arises from resonant second-order processes, where one photon is emitted and another one is absorbed.        

For the second resonance $p=q$, we observe that the effective Hamiltonian  is analogous to a two-photon Dicke Hamiltonian~\cite{compagno86,gerry89}
\begin{equation}
\label{eq:H2ph}
  \hat{H}_\text{2ph} = \frac{\alpha_N^2}{N}\left(\hat{a}_\text{L}^2\hat{\Upsilon}_{0,2}+
  \hat{a}_\text{L}^\dagger{}^2 \hat{\Upsilon}_{2,0}\right)  
\end{equation}
describing the transitions between the levels $q$ and $-q$ (compare to Tab.~\ref{tab:eff_Ham} of App.~\ref{sec:Effective_Hamiltonian}). Moreover,  we find a second contribution to this effective Hamiltonian that includes two-photon transitions, where one photon is emitted and one is absorbed in rough analogy to the origin of the Stark shift.

To derive an approximate solution for the second resonance we restrict ourselves for simplicity to the contribution corresponding to the two-photon Dicke Hamiltonian. In analogy to Refs.~\cite{PRR2021,kumar} we employ two constants of motion to find in App.~\ref{sec:Calculations_in_High-Gain_Regime} the expression
\begin{equation}
\label{eq:qhigh_nkl_zweite}
n_q(L)=n_0
\frac{1+\frac{n_0}{2N}}{\cos^2\left[\sqrt{\frac{n_0}{N}\left(\frac{n_0}{N}+2\right)}\frac{\alpha_N L}{2L_g}\right]+\frac{n_0}{2N}}
\end{equation}
for the mean photon number within a semi-classical approximation~\footnote{For a small interaction length $L$ we find the asymptotic behavior
$n\cong n_0 \left[1+n_0\left(\alpha_N L/L_g\right)^2/(2N)\right]$. Hence, a linear analysis (often used in FEL theory) is not sufficient to obtain this non-linear short-time behavior.}.  Moreover, we compute in App.~\ref{sec:Calculations_in_High-Gain_Regime} a numerical solution in rough analogy to the procedure for the fundamental resonance~\cite{walls70}. 

In the bottom panel of  Fig.~\ref{fig:n_q} we have drawn the mean photon number for $p=q$ as a function of the undulator length $L$. We observe that $n_q$  shows  an oscillatory behavior, with at most two emitted photons per electron. Compared to the solutions corresponding to the simplified model with the two-photon Dicke Hamiltonian, the curve emerging from the simulation of the full effective Hamiltonian of second order has a suppressed maximum which occurs after a slightly higher interaction length.

\begin{figure}
 \centering
  \includegraphics[scale=1]{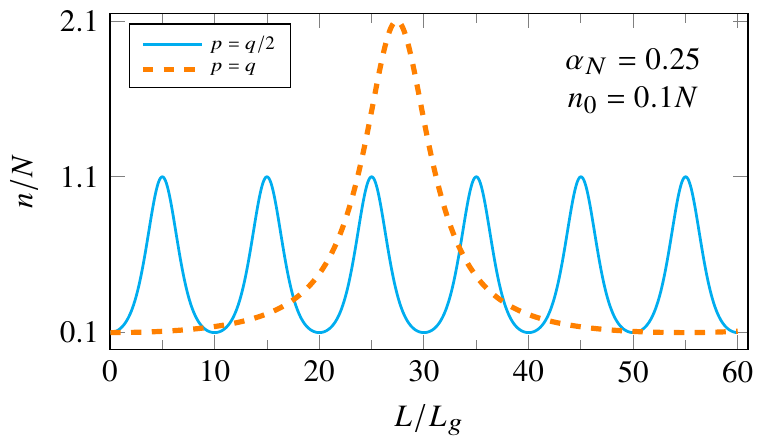}
 \caption{Mean photon number $n$ of a seeded high-gain FEL in the quantum regime divided by the number $N$ of electrons as a function of the undulator length $L$ in units of the gain length $L_g$. We compare the curves corresponding to the two analytical expressions Eqs.~\eqref{eq:nkl_erste} and~\eqref{eq:qhigh_nkl_zweite}, where the electrons start at (i) the first resonant momentum $p=q/2$ (blue line) and (ii) the second resonance $p=q$ (orange, dashed line).
We have chosen the values $n_0=0.1 N$ and $\alpha_N=0.25$ for the initial photon number and the quantum parameter, respectively. The second resonance leads to maximally two   emitted photons per electron compared to only one for the first resonance. However, for $p=q$ the growth of the photon number is much slower and the maximum occurs at a much higher interaction length compared to $p=q/2$. Hence, we deduce that the first resonance is more advantageous for the realization of a high-gain Quantum FEL than higher resonances.
}
 \label{fig:nkl}
\end{figure}

Similar to the low-gain regime, the maximum photon number increases for higher resonances, but at the same time the growth of the photon number becomes slower.
We identify this effect directly in the analytical results. For the second resonance the maximum photon number $n_\text{max}^q=n_0+2N$ occurs at the length $L_\text{max}^q$ while the corresponding maximum $n_\text{max}^{q/2}=n_0+N$ for $p=q/2$ is reached at $L_\text{max}^{q/2}$. With the help of Eqs.~\eqref{eq:nkl_erste} and~\eqref{eq:qhigh_nkl_zweite} we obtain the relation
\begin{equation}
\label{eq:Lmax_rel} 
\frac{L_\text{max}^q}{L_\text{max}^{q/2}}=\frac{1}{\alpha_N}\frac{\pi}{2\ln{\left(\sqrt{\frac{N}{n_0}}\right)}\sqrt{\frac{n_0}{N}\left(\frac{n_0}{N}+2\right)}}\,.
\end{equation}
%
Due to the scaling with $1/\alpha_N \gg 1$, the maximum for $p=q$ is shifted to the right compared to $p=q/2$. We visualize this behavior in Fig.~\ref{fig:nkl}, where we have drawn the mean photon numbers corresponding to these two resonances  both as functions of the undulator length~\footnote{Note that the results in the high-gain regime, Eqs.~\eqref{eq:nkl_erste} and~\eqref{eq:qhigh_nkl_zweite}, reduce to their respective low-gain counterparts in Eq.~\eqref{eq:dn_low} in the asymptotic limit $\delta n \sim N \ll n_0$.}. 
We derive from Eq.~\eqref{eq:Lmax_rel} with $n_0=0.1 N$ that $L_\text{max}^{q} \lesssim L_\text{max}^{q/2}$ only for $\alpha_N \gtrsim 3$ which is outside the quantum regime for which we require a small value of $\alpha_N$.

\section{Conclusions}

The quantum regime of the FEL emerges for high values of the quantum mechanical recoil, that is small wavelengths. Optical undulators are key~\cite{boni05,boni17} to  achieve such parameters experimentally. The requirements on power and pulse length of such a `pump laser'~\cite{steiniger} pose hard experimental challenges, already for the lowest-order~\cite{PRR2021} momentum resonance $p=q/2$. In addition, the combined influence of space charge and spontaneous emission limits the maximally possible interaction length~\cite{debus}.       
In this paper we demonstrated that higher-order resonant transitions require even larger undulator lengths due to the suppression of multiphoton transitions in the quantum regime.
As a consequence, the first resonance is favorable compared to the higher-order ones.


Moreover, we calculated multiphoton corrections to the deep quantum regime at $p=q/2$ in the low-gain~\cite{NJP2015} and for the first time also in the high-gain regime. Besides multiphoton processes, space charge and spontaneous emission can destroy the Quantum FEL dynamics~\cite{debus}. Only recently~\cite{schaap2022}, space-charge effects were studied in detail in a semic-classical phase-space model. In the next steps, one could combine all mentioned effects in a more complete Quantum FEL theory to specify more accurately parameter regimes, where an experimental relaization becomes possible.

\begin{acknowledgements}
We 
thank W.~P. Schleich, R.~Sauerbrey, C.~M. Carmesin, A.~Debus, and K. Steiniger 
for many exciting discussions.  
\end{acknowledgements}

\begin{appendix}
\section{Effective Hamiltonian}
\label{sec:Effective_Hamiltonian}

\begin{table*}
\caption{Effective Hamiltonian: We present for different resonant momenta the contributions of the asymptotic expansion of  $\hat{H}_\text{eff}$  in orders  of $\alpha_n$ in the low-gain regime and in orders of $\varepsilon$ in the high-gain regime, respectively. For the first and the second resonance, $p=q/2$ and $p=q$, we give the effective Hamiltonian up to third order. Moreover, we have calculated the fourth-order contribution in the case of $p=q$. For the third resonance $p=3q/2$ we have restricted ourselves to the low-gain regime.}
\begin{tabular}{lcc}
   \toprule
   & low gain: $ \hat{H}_\text{eff}\cong$ &  high gain: $\hat{H}_\text{eff}\cong$ \\
   \midrule
   \begin{tabular}{l}
      $\displaystyle p=\frac{q}{2}$ 
   \end{tabular}
   &
    \begin{tabular}{c}
      \begin{math}\displaystyle
        \alpha_n\left[\hat{\sigma}_{1,0}+\hat{\sigma}_{0,1}\vphantom{\hat{a}_{\text{L}}^\dagger \hat{\Upsilon}_{0,1}}\right]
        \end{math}
        \\
        \begin{math}\displaystyle
             + \alpha_n^2\left[
              -\frac{1}{2}\left(\hat{\sigma}_{0,0}+\hat{\sigma}_{1,1}\right)
              +\sum\limits_{\mu\neq0,1} \frac{\hat{\sigma}_{\mu,\mu}}{2\mu(\mu-1)}\right]
        \end{math}
        \\
        \begin{math}\displaystyle
            -\frac{\alpha_n^3}{4}
            \left[\hat{\sigma}_{0,1}+\hat{\sigma}_{1,0}-\hat{\sigma}_{-1,2}-\hat{\sigma}_{2,-1}   
            \right] \vphantom{\Bigg[}
        \end{math}
        \\
        \begin{math}\displaystyle
           \vphantom{\Bigg[}
        \end{math}
   \end{tabular}
    &
    \begin{tabular}{c}
       \begin{math} \displaystyle  
             \varepsilon\left[\hat{a}_{\text{L}} \hat{\Upsilon}_{1,0} 
             + \hat{a}_{\text{L}}^\dagger \hat{\Upsilon}_{0,1}  \right]
        \end{math}
        \\
        \begin{math} \displaystyle  
             +\frac{\varepsilon^2}{2}\left[\left(\hat{n}+1\right)
             \sum\limits_{\mu\neq 0}\frac{1}{\mu}
             \left(\hat{\Upsilon}_{\mu+1,\mu+1}-\hat{\Upsilon}_{\mu,\mu} \right)-
             \sum\limits_{\mu\neq 0}\frac{1}{\mu}\hat{\Upsilon}_{\mu+1,\mu}\hat{\Upsilon}_{\mu,\mu+1}\right]
        \end{math}
        \\
         \begin{math} \displaystyle  
            + \frac{\varepsilon^3}{4} \!\Bigg[
             \hat{a}_{\text{L}}\!\!\!\!\!\sum\limits_{\mu \neq -1,0}\!\!\frac{\hat{\Upsilon}_{2\mu +2,2\mu +1}
             \hat{\Upsilon}_{\mu,\mu+2}}{\mu(\mu+1)(2\mu+1)}
             \!+\!\frac{3\hat{a}_{\text{L}}}{2}\left(\hat{\Upsilon}_{0,-1}\hat{\Upsilon}_{-1,1}-\!\hat{\Upsilon}_{0,2}\hat{\Upsilon}_{2,1}\right)
        \end{math}
        \\
        \begin{math} \displaystyle 
           \phantom{+ \frac{\varepsilon^3}{4}} -\left(\sum\limits_{\mu\neq 0}\frac{\hat{\Upsilon}_{\mu+1,\mu+1}- \!\hat{\Upsilon}_{\mu,\mu}}{2\mu^2}
             \! +\!\hat{n} \!+\!\frac{1}{2}\right)
            \hat{a}_{\text{L}}\hat{\Upsilon}_{0,1}
             +\hat{a}_{\text{L}}^3\hat{\Upsilon}_{-1,2}+\text{h.c.}\Bigg]
        \end{math}        
    \end{tabular}
    \\
    \midrule
   \begin{tabular}{l}
      $\displaystyle p=q$ 
   \end{tabular}
    &
    \begin{tabular}{c}
        \begin{math}\displaystyle
          \alpha_n^2\left[
          \hat{\sigma}_{0,2}+\hat{\sigma}_{2,0}
          +\sum\limits_{\mu} \frac{2\,\hat{\sigma}_{\mu,\mu}}{(2\mu-3)(2\mu-1)}\right]
        \end{math}
        \\
         \begin{math}\displaystyle
            + \, \alpha_n^4\Bigg[-\frac{16}{9}(\hat{\sigma}_{2,0}+\hat{\sigma}_{0,2})
            +\frac{1}{36}(\hat{\sigma}_{3,-1}+\hat{\sigma}_{-1,3})
        \end{math}
        \\
        \begin{math}\displaystyle
            -\sum\limits_{\mu}\frac{\hat{\sigma}_{\mu+1,\mu+1}-\hat{\sigma}_{\mu,\mu}}{8(\mu-1/2)^3 \left\lbrace(\mu-1/2)^2-1\right\rbrace^2}
        \end{math}
        \\
        \begin{math}\displaystyle
            +\!\!\sum\limits_{\mu\neq 0}\frac{\hat{\sigma}_{\mu+1,\mu+1}+\hat{\sigma}_{\mu,\mu}}{64\mu (\mu^2-1/4)^2}
            \Bigg]
        \end{math}
    \end{tabular}
     &
    \begin{tabular}{c}
        \begin{math}\displaystyle
            \varepsilon^2 \Bigg[\hat{a}_{\text{L}}{}^2\hat{\Upsilon}_{0,2}+ \text{ h.c.} +\hat{n}\!\sum\limits_\mu\frac{\hat{\Upsilon}_{\mu+1,\mu+1}}{2\mu-1} - \left(\hat{n}+1\right)\sum\limits_\mu\frac{\hat{\Upsilon}_{\mu,\mu}}{2\mu-1}
        \end{math}
        \\
        \begin{math}\displaystyle
        \phantom{ \varepsilon^2 \Bigg[}
           +\!\sum\limits_\mu\!\frac{\hat{\Upsilon}_{\mu+1,\mu+1}+\hat{\Upsilon}_{\mu,\mu}\!-\!\hat{\Upsilon}_{\mu+1,\mu}\hat{\Upsilon}_{\mu,\mu   +1}-\hat{\Upsilon}_{\mu,\mu+1}\hat{\Upsilon}_{\mu+1,\mu}}{4\mu -2}\!\Bigg]
        \end{math} 
        \\
        \begin{math}\displaystyle
           \phantom{
            -\sum\limits_{\mu}\frac{\hat{\sigma}_{\mu+1,\mu+1}}{\left\lbrace(\mu)^2\right\rbrace^2}
            }
        \end{math}
        \\
        \begin{math}\displaystyle
            \phantom{
            \sum\limits_{\mu\neq 0}\frac{\hat{\sigma}_{\mu+1,\mu+1}}{ (\mu^2)^2}
            \Bigg]
            }
        \end{math}
    \end{tabular}
    \\
    \midrule   
   \begin{tabular}{l}
      $\displaystyle p=\frac{3q}{2}$
    \end{tabular}
    &
    \begin{tabular}{c}
        $ \displaystyle \alpha_n\left[\hat{\sigma}_{1,2}+\hat{\sigma}_{2,1}\right]$
        \\
        \begin{math}\displaystyle
            +\alpha_n^2\left[
            -\frac{1}{2}\left(\hat{\sigma}_{1,1}+\hat{\sigma}_{2,2}\right)
            +\sum\limits_{\mu\neq 1,2} \frac{\hat{\sigma}_{\mu,\mu}}{2(\mu-1)(\mu-2)}\right]
        \end{math}
        \\
         \begin{math}\displaystyle
            +\frac{\alpha_n^3}{4}\left[
            \hat{\sigma}_{0,3}+\hat{\sigma}_{3,0}-\hat{\sigma}_{1,2}-\hat{\sigma}_{2,1}\right]
        \end{math}
    \end{tabular}
    &   
  \\   
    \bottomrule
  \end{tabular}
  \label{tab:eff_Ham} 
\end{table*} 

We start with the dimensionless Hamiltonian in the high-gain regime~\cite{PRA2019}
 \begin{equation}
    \label{eq:H_many}
        \hat{H}\!\equiv \! \varepsilon\!\!\sum\limits_\mu \!\left(\e{\I 2\tau\left[\frac{p}{q}-\left(\mu +\frac{1}{2}\right)\right]}\!\hat{a}_{\text{L}}\hat{\Upsilon}_{\mu,\mu+1}\!+ \text{ h.c.}\right)
      \end{equation}
in the interaction picture with the dimensionless time variable $\tau \equiv \omega_\text{r}t$. To obtain the single-electron and semi-classical Hamiltonian for a low-gain FEL, we simply have to replace the collective operators $\hat{\Upsilon}_{\mu,\nu}$ by their single-particle counterparts $\hat{\sigma}_{\mu,\nu}$ and approximate $\hat{a}_\text{L}\approx \hat{a}_\text{L}^\dagger\approx \sqrt{n}\approx \text{const}$. We note that the commutation relation
\begin{equation}
\label{eq:Upsilon_comm}
\left[\hat{\Upsilon}_{\mu,\nu},\hat{\Upsilon}_{\rho,\sigma}\right]
=\updelta_{\nu,\rho}\hat{\Upsilon}_{\mu,\sigma}
-\updelta_{\sigma,\mu}\hat{\Upsilon}_{\rho,\nu}
\end{equation}
for the jump operators is the same for the collective model as in the single-electron limit. However, the properties of products of these operators differ~\cite{PRA2019}.  

The asymptotic method of averaging~\cite{bogoliubov,higher,peter} is suitable for a Hamiltonian $\hat{H}$ which can be represented as a Fourier series in terms of the phase $ \tau$ and its integer multiples. We separate slow and rapid dynamics in the state vector %
$\ket{\Psi(\tau)}\equiv \exp[{-\hat{F}(\tau)]}\ket{\Phi(\tau)}$, %
where $\hat{F}$ describes the rapidly varying part, while $\ket{\Phi}$ gives the slowly-varying part. With the help of this ansatz we derive the effective Hamiltonian~\cite{higher}
\begin{equation}\label{eq:H_eff_allg}
\hat{H}_\text{eff}=\sum\limits_{j=0}^\infty\frac{1}{(j+1)!}\left[\hat{F},\I\frac{\D\hat{F}}{\D\tau}\right]_j+\sum\limits_{j=0}^\infty\frac{1}{j!}\left[\hat{F},\hat{H}\right]_j
\end{equation}
of the Schr{\"o}dinger equation for $\ket{\Phi}$, where the subscript $j$ indicates a $j$ times nested commutator.

We proceed by asymptotically expanding $\hat{H}_\text{eff}$ and $\hat{F}$ in powers of $\alpha_n$ -- or in powers of $\varepsilon\equiv g/\omega_\text{r}$ in the high-gain regime. In each order of this expansion we have to ensure that the effective Hamiltonian is independent of time, that is $\hat{H}_\text{eff}\neq \hat{H}_\text{eff}(\tau)$. Hereby, we avoid secular contributions which otherwise lead to unphysically growing terms~\cite{nayfeh}. The dynamics dictated by $\hat{H}_\text{eff}$ can then to be solved non-perturbatively. In contrast, we can rely on perturbation theory for the rapidly-varying dynamics since here the secular terms are excluded by construction.

Depending on the specific initial momentum $p=\nu q/2$ with integer $\nu$ we  obtain from Eq.~\eqref{eq:H_many} the explicit  expressions for the Fourier components  of $\hat{H}$.
By inserting these components into $\hat{H}_{\text{eff}}$ from Eq.~\eqref{eq:H_eff_allg} and calculating the occurring commutators we finally obtain the effective Hamiltonian for low and high gain and for different resonances. We have listed the   
explicit expressions in Tab.~\ref{tab:eff_Ham}.

\section{Population of Momentum Levels}
\label{sec:Population_of_Momentum_Levels}

In this appendix we discuss the population probabilities of the momentum levels for an electron in a low-gain FEL resulting from the asymptotic method of averaging. For the first resonance $p=q/2$ we refer to Ref.~\cite{NJP2015}, where the population probabilities for the momentum levels are listed up to third order in $\alpha_n$ for the frequency and up to  second order  for the amplitude. In the following we consider the second and the third resonance. 
  
\subsection{Second resonance}

The initial state of an electron for the second resonance is given by the momentum eigenstate   $\ket{\Psi(0)}=\ket{p}$ with $p=q$. However, due to the transformation from $\ket{\Psi}$ to 
$\ket{\Phi}$ we calculate the transformed initial state 
$\ket{\Phi(0)}=\exp[{\hat{F}(0)]}\ket{\Psi(0)}$ perturbatively up to second order of $\alpha_n$.


We expand the state $\ket{\Phi}$  in the discretized momentum basis with probability amplitudes $ \braket{p-\mu q|\Phi(\tau)}$. The Schr{\"o}dinger equation corresponding to the effective Hamiltonian from Tab.~\ref{tab:eff_Ham}  then translates to a system of linear  differential equations which we easily solve with respect to the initial conditions for $\ket{\Phi}$.

Then, we transform the result for  $\ket{\Phi}$ back to the original state 
$\ket{\Psi}$ via the relation $\ket{\Psi(\tau)}=\exp[{-\hat{F}(\tau)]}\ket{\Phi(\tau)}$, and again restrict ourselver to terms up to second order of $\alpha_n$. Finally, we calculate the probabilities %
$P_{p-\mu q}(\tau)\equiv \left|\braket{p-\mu q|\Psi(\tau)}\right|^2$ %
for the population of the momentum levels up to the order $\alpha_n^2$ in amplitude and $\alpha_n^4$ in frequency.

By this procedure, we find the explicit expressions
\allowdisplaybreaks
\begin{align}
P_{2q}(\tau)&=\! \frac{\alpha_n^2}{9}\!\left(\cos^2{\xi_1 \tau}\!+\!\cos^2{\xi_2 \tau} \!-\!2\cos{\xi_1\tau} 
\cos{\xi_2\tau}
\cos{\xi_3 \tau}\right)\nonumber \\ \nonumber
P_q(\tau)&= \cos^2{\xi_1 \tau} +2\alpha_n^2\cos{\xi_1\tau} \\ \nonumber
& \ \ \ \ \ \times\left(-\frac{10}{9}\cos{\xi_1\tau}
+\cos{\xi_4\tau}  +\frac{1}{9}\cos{\xi_2\tau}
\cos{\xi_3\tau}\right)\\
P_{0}(\tau)&=2\alpha_n^2\left\{1-\cos{\left[\left(\xi_1+\xi_4\right)\tau\right]}\right\}\nonumber \\ \nonumber
P_{-q}(\tau)&= \sin^2{\xi_1\tau}
+2\alpha_n^2\sin{\xi_1\tau}\\ \nonumber
& \ \ \ \ \ \times \left(-\frac{10}{9}\sin{\xi_1\tau}
-\sin{\xi_4\tau}+\frac{1}{9}\sin{\xi_2\tau}
\cos{\xi_3\tau}\right)
 \ \ \ \ \ \ 
\\
P_{-2q}(\tau)&= \frac{\alpha_n^2}{9}\!\left(\sin^2{\xi_1\tau}+\sin^2{\xi_2\tau}-2\sin{\xi_1\tau} 
\sin{\xi_2\tau}
\cos{\xi_3\tau}\right) \nonumber
\end{align}
with 
\allowdisplaybreaks
\begin{align}
  \xi_1 & \equiv  \alpha_n^2\left(1-\frac{16\alpha_n^2}{9}\right)
  \nonumber\\
  \xi_2 &\equiv \frac{\alpha_n^4}{36}\sqrt{1+\left(\frac{124}{125}\right)^2} \nonumber\\ 
  \xi_3 &\equiv 3- \frac{8\alpha_n^2}{15}\left(1-\frac{16\alpha_n^2}{5}\right) \nonumber\\
  \xi_4 &\equiv 1+\frac{8\alpha_n^2}{3}\left(1-7\left(\frac{8\alpha_n}{15}\right)^2\right)\,. \nonumber
\end{align}
We note that the sum over these probabilities equals unity.

\subsection{Third resonance}

For the third resonance, $p=3q/2$, we neglect the amplitude corrections and assume that  $\ket{\Psi}\approx\ket{\Phi}$. With the help of the effective Hamiltonian in Tab.~\ref{tab:eff_Ham}
we obtain the probabilities 
\begin{align}
\label{eq:app_P3q}
P_{3q/2}(\tau) = \cos^2{\left(\frac{\alpha_n^3\tau}{4}\right)} \ \   \text{and} \ \   
P_{-3q/2} (\tau)= \sin^2{\left(\frac{\alpha_n^3\tau}{4}\right)}
\end{align}
for the population of the momentum levels $3q/2$ and $-3q/2$, respectively.

\section{Calculations in High-Gain Regime} 
\label{sec:Calculations_in_High-Gain_Regime}

We calculate the time evolution of the mean photon number for a high-gain FEL in the quantum regime at the second resonance. For that we employ (i) an analytical approximation and (ii) a numerical simulation. 
 
\subsection{Analytical approximation}
\label{sec:Analytic approximation}

The momentum jump operators appearing in the two-photon Dicke Hamiltonian $\hat{H}_{2\text{ph}}$ from Eq.~\eqref{eq:H2ph} can be treated analogously to ladder operators of angular momenta.  
For simplicity, we employ the Schwinger representation of angular momentum~\cite{schwinger} by introducing the bosonic annihilation and creation operators, $\hat{b}_s$ and $\hat{b}_s^\dagger$, respectively for two modes $s=0,2$. We then identify the relations $\hat{\Upsilon}_{0,2}\equiv \hat{b}_0^\dagger \hat{b}_2$ and
$\hat{\Upsilon}_{2,0}\equiv \hat{b}_2^\dagger \hat{b}_0$. Hence, we obtain the Hamiltonian
\begin{equation}
\hat{H}_{2\text{ph}}=\varepsilon^2 \left(\hat{a}_{\text{L}}{}^2\hat{b}_0^\dagger\hat{b}_2+\hat{a}_{\text{L}}^\dagger{}^2\hat{b}_2^\dagger\hat{b}_0\right)
\end{equation}
from which we derive via the Heisenberg equations of motion the two constants of motion 
   $\hat{A}  \equiv \hat{N}_0 + \hat{N}_2=\text{const}$, and 
   $\hat{B}  \equiv 2\hat{N}_0 + \hat{n}=\text{const}$ %
with $\hat{N}_k\equiv \hat{b}_k^\dagger \hat{b}_k $ and $\hat{n}\equiv\hat{a}_{\text{L}}^\dagger\hat{a}_{\text{L}} $~\cite{kumar}.

In the following we approximate the bosonic operators as classical but dynamical changing variables. The Hamiltonian equation of motion for a dynamical quantity $f$ 
then reads
\begin{equation}
\frac{\D f}{\D \tau}=\left\{f,H_{2\text{ph}}\right\}\equiv -\I \!\! \! \sum\limits_{s=0,2,\text{L}}\left(\frac{\partial f}{\partial b_s}\frac{\partial H_{2\text{ph}}}{\partial b_s^*}-\frac{\partial f}{\partial b_s^*}\frac{\partial H_{2\text{ph}}}{\partial b_s}\right),
\end{equation}
where we have defined the Poisson brackets for the complex amplitudes $b_0$, $b_2$ und $b_{\text{L}}\equiv a_{\text{L}}$ of three harmonic oscillators. This semi-classical approximation neglects contributions that are responsible for spontaneous emission and thus we deduce that our approximation works for a seeded FEL, but breaks down for self-amplified spontaneous emission (SASE).

For the time evolution of the photon number $n\equiv |a_{\text{L}}|^2$ we obtain the second-order differential equation
\begin{equation}
\label{eq:ddot}
\ddot{n}=4\varepsilon^4\left[4n N_0 N_2+n_0^2 (N_0-N_2)\right]
\end{equation}  
with $N_s\equiv |b_s|^2 $. We assume that the two constants of motion, $\hat{A}$ and $\hat{B}$, are described by their initial expectations values, that is $A=N $ and $B=2N+n_0$, respectively. With the help of these relations we eliminate $N_0$ and $N_2$ in Eq.~\eqref{eq:ddot} and obtain a closed equation for $n$. After integrating twice with respect to time $\tau$ we observe
\begin{equation}
2\alpha_N^2\tau=\int\limits_{n_0/N}^{n/N}\frac{\D\xi}{\xi \sqrt{\left(\xi-\frac{n_0}{N}\right)
\left(2+\frac{n_0}{N}-\xi\right)}}
\end{equation}
which can be solved analytically. Finally, we arrive at the expression in Eq.~\eqref{eq:qhigh_nkl_zweite} for the evolution of the photon number $n=n(L)$, where we have introduced the interaction length $L$ via the relation $\alpha_N\tau=L/(2L_g)$~\cite{PRA2019}. 

\subsection{Numerical simulation}

To find a numerical solution for the dynamics dictated by the effective Hamiltonian for $p=q$  we first consider the contribution corresponding to the two-photon Dicke Hamiltonian $\hat{H}_{2\text{ph}}$. Similarly to Ref.~\cite{PRR2021} we notice the analogy of the jump operators to angular momentum, that is $ \hat{J}_+ = \hat{\Upsilon}_{0,2}$, $\hat{J}_- = \hat{\Upsilon}_{2,0}$, and
$\hat{J}_z=(\hat{\Upsilon}_{0,0}-\hat{\Upsilon}_{2,2})/2$. By applying the ladder operators $\hat{J}_\pm$ on the state $\ket{r,m}$ we obtain the relation~\cite{ct}
\begin{equation}
\label{eq:app_Jpm}
\hat{J}_\pm \ket{r,m}=\sqrt{(r\pm m+1)(r\mp m)}\ket{r,m\pm 1}\,,
\end{equation}
where $r$ and $m$ correspond to the quantum numbers of total angular momentum and its $z$-component, respectively.

In this description, the initial state of the electrons is given by %
$\ket{N/2,N/2}=\ket{p,p,...,p}$. %
In this case, only superpositions of the following states 
\begin{equation}
\label{eq:app_mu}
\ket{\mu}\equiv \ket{n_0+2\mu}\ket{N/2,N/2-\mu}
\end{equation}
can be populated by $\hat{H}_{2\text{ph}}$, if we assume that the laser field starts from a Fock state with $n_0$ photons~\cite{walls70}. The quantum number $\mu$ runs from $0$ to $N$, due to 
$-r \leq m \leq r$ with $r=N/2$.

We note that the second contribution $\hat{\Delta}\equiv \hat{H}_\text{eff}-\hat{H}_{2\text{ph}}$ to the effective Hamiltonian (compare to Tab.~\ref{tab:eff_Ham}) includes operators outside this angular momentum algebra. To proceed, we write the electron part of the state in Eq.~\eqref{eq:app_mu} in the  form
\begin{equation}
\ket{N/2,N/2-\mu}=\frac{1}{\sqrt{\mu !}}\sqrt{\frac{(N-\mu)!}{N!}}\,\,\hat{J}_-^\mu\,  \ket{N/2,N/2}
\end{equation}
which follows from Eq.~\eqref{eq:app_Jpm}. With the help of this relation and the commutation relation for the jump operators in Eq.~\eqref{eq:Upsilon_comm} we calculate the action of $\hat{\Delta}$ on the state $\ket{\mu}$ and find that it is an eigenstate of  $\hat{\Delta}$. Hence, we still can rely on the formalism for $\hat{H}_{2\text{ph}}$ for the full effective Hamiltonian since $\hat{\Delta}$ reproduces only states in the form of Eq.~\eqref{eq:app_mu}.

After expanding the quantum state $\ket{\Psi}$ of the total system in terms of the basis states
$\ket{\mu}$, and applying the Schrödinger equation with the effective Hamiltonian for an initial momentum $p$, we finally obtain the equation of motion  
\begin{equation}
\label{eq:cdot}
\I\frac{\D c_\mu(L)}{\D(L/L_g)}=a_p(\mu)c_{\mu-1}(L)+a(\mu+1)c_{\mu+1}(L)+d_p(\mu)c_\mu(L)
\end{equation}
for the expansion coefficients $c_\mu\equiv \braket{\mu|\Psi} $.
For $p=q$, the off-diagonal terms
\begin{subequations}
\begin{equation}
a_q(\mu)\equiv \frac{\alpha_N}{2}\sqrt{(n_0+2\mu-1)(n_0+2\mu)}\sqrt{\frac{\mu}{N}}\sqrt{1-\frac{\mu-1}{N}}
\end{equation}
emerge from the two-photon Dicke Hamiltonian $\hat{H}_{2\text{ph}}$ and
\begin{align}
d_q(\mu)=\alpha_N\left[\frac{2}{3}\mu\left(1-\frac{1}{N}\right)+\frac{1}{3}n_0+\frac{1}{2}\right]
\end{align}
\end{subequations}
represents the additional diagonal contributions arising from $\hat{\Delta}$. Similarly to App.~\ref{sec:Analytic approximation}, we transformed from $\tau$ to $L$. The probability amplitudes $c_\mu$ contain all information of the quantum state of the system and after computing them numerically by diagonalizing a $(N+1)\times (N+1)$ tri-diagonal matrix we are able to evaluate any expectation value.

Analogously, we find for the resonance $p=q/2$ a dynamical equation of the same form as Eq.~\eqref{eq:cdot} using the corresponding effective Hamiltonian from Tab.~\ref{tab:eff_Ham} up to third order. In this case, the ladder operators of angular momentum  are given by $\hat{\Upsilon}_{1,0}$ and $\hat{\Upsilon}_{0,1}$. We obtain the expressions
\begin{subequations}
\begin{equation}
\label{eq:aq2}
a_{q/2}(\mu)\equiv
\frac{1}{2}\!\left[1\!-\!\frac{\alpha_N^2}{8}\!\left(1+2\frac{n_0+1}{N}\right)\!\right]\sqrt{\mu(n_0+\mu)}\sqrt{1-\frac{\mu-1}{N}}
\end{equation}
and
\begin{equation}
\label{eq:dq2}
d_{q/2}(\mu)\equiv -\frac{\alpha_N}{4}\left[n_0+\mu\left(1+\frac{1}{N}\right)\right]\,
\end{equation}
\end{subequations}
for the off-diagonal and diagonal terms in the differential equation.

\end{appendix}

%


\begin{thebibliography}{40}%
\makeatletter
\providecommand \@ifxundefined [1]{%
 \@ifx{#1\undefined}
}%
\providecommand \@ifnum [1]{%
 \ifnum #1\expandafter \@firstoftwo
 \else \expandafter \@secondoftwo
 \fi
}%
\providecommand \@ifx [1]{%
 \ifx #1\expandafter \@firstoftwo
 \else \expandafter \@secondoftwo
 \fi
}%
\providecommand \natexlab [1]{#1}%
\providecommand \enquote  [1]{``#1''}%
\providecommand \bibnamefont  [1]{#1}%
\providecommand \bibfnamefont [1]{#1}%
\providecommand \citenamefont [1]{#1}%
\providecommand \href@noop [0]{\@secondoftwo}%
\providecommand \href [0]{\begingroup \@sanitize@url \@href}%
\providecommand \@href[1]{\@@startlink{#1}\@@href}%
\providecommand \@@href[1]{\endgroup#1\@@endlink}%
\providecommand \@sanitize@url [0]{\catcode `\\12\catcode `\$12\catcode
  `\&12\catcode `\#12\catcode `\^12\catcode `\_12\catcode `\%12\relax}%
\providecommand \@@startlink[1]{}%
\providecommand \@@endlink[0]{}%
\providecommand \url  [0]{\begingroup\@sanitize@url \@url }%
\providecommand \@url [1]{\endgroup\@href {#1}{\urlprefix }}%
\providecommand \urlprefix  [0]{URL }%
\providecommand \Eprint [0]{\href }%
\providecommand \doibase [0]{https://doi.org/}%
\providecommand \selectlanguage [0]{\@gobble}%
\providecommand \bibinfo  [0]{\@secondoftwo}%
\providecommand \bibfield  [0]{\@secondoftwo}%
\providecommand \translation [1]{[#1]}%
\providecommand \BibitemOpen [0]{}%
\providecommand \bibitemStop [0]{}%
\providecommand \bibitemNoStop [0]{.\EOS\space}%
\providecommand \EOS [0]{\spacefactor3000\relax}%
\providecommand \BibitemShut  [1]{\csname bibitem#1\endcsname}%
\let\auto@bib@innerbib\@empty
\bibitem [{\citenamefont {Schroeder}\ \emph {et~al.}(2001)\citenamefont
  {Schroeder}, \citenamefont {Pellegrini},\ and\ \citenamefont
  {Chen}}]{schroeder}%
  \BibitemOpen
  \bibfield  {author} {\bibinfo {author} {\bibfnamefont {C.~B.}\ \bibnamefont
  {Schroeder}}, \bibinfo {author} {\bibfnamefont {C.}~\bibnamefont
  {Pellegrini}},\ and\ \bibinfo {author} {\bibfnamefont {P.}~\bibnamefont
  {Chen}},\ }\bibfield  {title} {\bibinfo {title} {Quantum effects in high-gain
  free-electron lasers},\ }\href {https://doi.org/10.1103/PhysRevE.64.056502}
  {\bibfield  {journal} {\bibinfo  {journal} {Phys. Rev. E}\ }\textbf {\bibinfo
  {volume} {64}},\ \bibinfo {pages} {056502} (\bibinfo {year}
  {2001})}\BibitemShut {NoStop}%
\bibitem [{\citenamefont {Piovella}\ \emph {et~al.}(2008)\citenamefont
  {Piovella}, \citenamefont {Cola}, \citenamefont {Volpe}, \citenamefont
  {Schiavi},\ and\ \citenamefont {Bonifacio}}]{pio_prl}%
  \BibitemOpen
  \bibfield  {author} {\bibinfo {author} {\bibfnamefont {N.}~\bibnamefont
  {Piovella}}, \bibinfo {author} {\bibfnamefont {M.~M.}\ \bibnamefont {Cola}},
  \bibinfo {author} {\bibfnamefont {L.}~\bibnamefont {Volpe}}, \bibinfo
  {author} {\bibfnamefont {A.}~\bibnamefont {Schiavi}},\ and\ \bibinfo {author}
  {\bibfnamefont {R.}~\bibnamefont {Bonifacio}},\ }\bibfield  {title} {\bibinfo
  {title} {Three-dimensional {W}igner-function description of the quantum
  free-electron laser},\ }\href
  {https://doi.org/10.1103/PhysRevLett.100.044801} {\bibfield  {journal}
  {\bibinfo  {journal} {Phys. Rev. Lett.}\ }\textbf {\bibinfo {volume} {100}},\
  \bibinfo {pages} {044801} (\bibinfo {year} {2008})}\BibitemShut {NoStop}%
\bibitem [{\citenamefont {Bonifacio}\ \emph {et~al.}(2017)\citenamefont
  {Bonifacio}, \citenamefont {Fares}, \citenamefont {Ferrario}, \citenamefont
  {McNeil},\ and\ \citenamefont {Robb}}]{boni17}%
  \BibitemOpen
  \bibfield  {author} {\bibinfo {author} {\bibfnamefont {R.}~\bibnamefont
  {Bonifacio}}, \bibinfo {author} {\bibfnamefont {H.}~\bibnamefont {Fares}},
  \bibinfo {author} {\bibfnamefont {M.}~\bibnamefont {Ferrario}}, \bibinfo
  {author} {\bibfnamefont {B.~W.~J.}\ \bibnamefont {McNeil}},\ and\ \bibinfo
  {author} {\bibfnamefont {G.~R.~M.}\ \bibnamefont {Robb}},\ }\bibfield
  {title} {\bibinfo {title} {Design of sub-{A}ngstrom compact free-electron
  laser source},\ }\href {https://doi.org/10.1016/j.optcom.2016.07.007}
  {\bibfield  {journal} {\bibinfo  {journal} {Opt. Commun.}\ }\textbf {\bibinfo
  {volume} {382}},\ \bibinfo {pages} {58} (\bibinfo {year} {2017})}\BibitemShut
  {NoStop}%
\bibitem [{\citenamefont {Serbeto}\ \emph {et~al.}(2009)\citenamefont
  {Serbeto}, \citenamefont {Monteiro}, \citenamefont {Tsui},\ and\
  \citenamefont {Mendon\c{c}a}}]{serbeto09}%
  \BibitemOpen
  \bibfield  {author} {\bibinfo {author} {\bibfnamefont {A.}~\bibnamefont
  {Serbeto}}, \bibinfo {author} {\bibfnamefont {L.~F.}\ \bibnamefont
  {Monteiro}}, \bibinfo {author} {\bibfnamefont {K.~H.}\ \bibnamefont {Tsui}},\
  and\ \bibinfo {author} {\bibfnamefont {J.~T.}\ \bibnamefont {Mendon\c{c}a}},\
  }\bibfield  {title} {\bibinfo {title} {Quantum plasma fluid model for
  high-gain free-electron lasers},\ }\href
  {https://doi.org/10.1088/0741-3335/51/12/124024} {\bibfield  {journal}
  {\bibinfo  {journal} {Plasma Phys. Control. Fusion}\ }\textbf {\bibinfo
  {volume} {51}},\ \bibinfo {pages} {124024} (\bibinfo {year}
  {2009})}\BibitemShut {NoStop}%
\bibitem [{\citenamefont {Kling}\ \emph {et~al.}(2015)\citenamefont {Kling},
  \citenamefont {Giese}, \citenamefont {Endrich}, \citenamefont {Preiss},
  \citenamefont {Sauerbrey},\ and\ \citenamefont {Schleich}}]{NJP2015}%
  \BibitemOpen
  \bibfield  {author} {\bibinfo {author} {\bibfnamefont {P.}~\bibnamefont
  {Kling}}, \bibinfo {author} {\bibfnamefont {E.}~\bibnamefont {Giese}},
  \bibinfo {author} {\bibfnamefont {R.}~\bibnamefont {Endrich}}, \bibinfo
  {author} {\bibfnamefont {P.}~\bibnamefont {Preiss}}, \bibinfo {author}
  {\bibfnamefont {R.}~\bibnamefont {Sauerbrey}},\ and\ \bibinfo {author}
  {\bibfnamefont {W.~P.}\ \bibnamefont {Schleich}},\ }\bibfield  {title}
  {\bibinfo {title} {What defines the quantum regime of the free-electron
  laser?},\ }\href {https://doi.org/10.1088/1367-2630/17/12/123019} {\bibfield
  {journal} {\bibinfo  {journal} {New J. Phys.}\ }\textbf {\bibinfo {volume}
  {17}},\ \bibinfo {pages} {123019} (\bibinfo {year} {2015})}\BibitemShut
  {NoStop}%
\bibitem [{\citenamefont {Brown}\ \emph {et~al.}(2017)\citenamefont {Brown},
  \citenamefont {Henderson}, \citenamefont {Campbell},\ and\ \citenamefont
  {McNeil}}]{brown17}%
  \BibitemOpen
  \bibfield  {author} {\bibinfo {author} {\bibfnamefont {M.~S.}\ \bibnamefont
  {Brown}}, \bibinfo {author} {\bibfnamefont {J.~R.}\ \bibnamefont
  {Henderson}}, \bibinfo {author} {\bibfnamefont {L.~T.}\ \bibnamefont
  {Campbell}},\ and\ \bibinfo {author} {\bibfnamefont {B.~W.~J.}\ \bibnamefont
  {McNeil}},\ }\bibfield  {title} {\bibinfo {title} {An extended model of the
  quantum free-electron laser},\ }\href {https://doi.org/10.1364/OE.25.033429}
  {\bibfield  {journal} {\bibinfo  {journal} {Opt. Express}\ }\textbf {\bibinfo
  {volume} {25}},\ \bibinfo {pages} {33429} (\bibinfo {year}
  {2017})}\BibitemShut {NoStop}%
\bibitem [{\citenamefont {Schaap}\ \emph {et~al.}(2022)\citenamefont {Schaap},
  \citenamefont {Schouwenaars},\ and\ \citenamefont {Luiten}}]{schaap2022}%
  \BibitemOpen
  \bibfield  {author} {\bibinfo {author} {\bibfnamefont {B.~H.}\ \bibnamefont
  {Schaap}}, \bibinfo {author} {\bibfnamefont {S.}~\bibnamefont
  {Schouwenaars}},\ and\ \bibinfo {author} {\bibfnamefont {O.~J.}\ \bibnamefont
  {Luiten}},\ }\bibfield  {title} {\bibinfo {title} {A {R}aman quantum
  free-electron laser model},\ }\href {https://doi.org/10.1063/5.0106439}
  {\bibfield  {journal} {\bibinfo  {journal} {Phys. Plasmas}\ }\textbf
  {\bibinfo {volume} {29}},\ \bibinfo {pages} {113302} (\bibinfo {year}
  {2022})}\BibitemShut {NoStop}%
\bibitem [{\citenamefont {Bonifacio}\ \emph {et~al.}(2006)\citenamefont
  {Bonifacio}, \citenamefont {Piovella}, \citenamefont {Robb},\ and\
  \citenamefont {Schiavi}}]{boni06}%
  \BibitemOpen
  \bibfield  {author} {\bibinfo {author} {\bibfnamefont {R.}~\bibnamefont
  {Bonifacio}}, \bibinfo {author} {\bibfnamefont {N.}~\bibnamefont {Piovella}},
  \bibinfo {author} {\bibfnamefont {G.~R.~M.}\ \bibnamefont {Robb}},\ and\
  \bibinfo {author} {\bibfnamefont {A.}~\bibnamefont {Schiavi}},\ }\bibfield
  {title} {\bibinfo {title} {Quantum regime of free electron lasers starting
  from noise},\ }\href {https://doi.org/10.1103/PhysRevSTAB.9.090701}
  {\bibfield  {journal} {\bibinfo  {journal} {Phys. Rev. Spec. Top.--Accel.
  Beams}\ }\textbf {\bibinfo {volume} {9}},\ \bibinfo {pages} {090701}
  (\bibinfo {year} {2006})}\BibitemShut {NoStop}%
\bibitem [{\citenamefont {Kling}\ \emph {et~al.}(2021)\citenamefont {Kling},
  \citenamefont {Giese}, \citenamefont {Carmesin}, \citenamefont {Sauerbrey},\
  and\ \citenamefont {Schleich}}]{PRR2021}%
  \BibitemOpen
  \bibfield  {author} {\bibinfo {author} {\bibfnamefont {P.}~\bibnamefont
  {Kling}}, \bibinfo {author} {\bibfnamefont {E.}~\bibnamefont {Giese}},
  \bibinfo {author} {\bibfnamefont {C.~M.}\ \bibnamefont {Carmesin}}, \bibinfo
  {author} {\bibfnamefont {R.}~\bibnamefont {Sauerbrey}},\ and\ \bibinfo
  {author} {\bibfnamefont {W.~P.}\ \bibnamefont {Schleich}},\ }\bibfield
  {title} {\bibinfo {title} {High-gain quantum free-electron laser: Long-time
  dynamics and requirements},\ }\href
  {https://doi.org/10.1103/PhysRevResearch.3.033232} {\bibfield  {journal}
  {\bibinfo  {journal} {Phys. Rev. Research}\ }\textbf {\bibinfo {volume}
  {3}},\ \bibinfo {pages} {033232} (\bibinfo {year} {2021})}\BibitemShut
  {NoStop}%
\bibitem [{\citenamefont {Debus}\ \emph {et~al.}(2019)\citenamefont {Debus},
  \citenamefont {Steiniger}, \citenamefont {Kling}, \citenamefont {Carmesin},\
  and\ \citenamefont {Sauerbrey}}]{debus}%
  \BibitemOpen
  \bibfield  {author} {\bibinfo {author} {\bibfnamefont {A.}~\bibnamefont
  {Debus}}, \bibinfo {author} {\bibfnamefont {K.}~\bibnamefont {Steiniger}},
  \bibinfo {author} {\bibfnamefont {P.}~\bibnamefont {Kling}}, \bibinfo
  {author} {\bibfnamefont {C.~M.}\ \bibnamefont {Carmesin}},\ and\ \bibinfo
  {author} {\bibfnamefont {R.}~\bibnamefont {Sauerbrey}},\ }\bibfield  {title}
  {\bibinfo {title} {Realizing quantum free-electron lasers: a critical
  analysis of experimental challenges and theoretical limits},\ }\href
  {https://doi.org/10.1088/1402-4896/aaf951} {\bibfield  {journal} {\bibinfo
  {journal} {Phys. Scr.}\ }\textbf {\bibinfo {volume} {94}},\ \bibinfo {pages}
  {074001} (\bibinfo {year} {2019})}\BibitemShut {NoStop}%
\bibitem [{\citenamefont {Seddon}\ \emph {et~al.}(2017)\citenamefont {Seddon},
  \citenamefont {Clarke}, \citenamefont {Dunning}, \citenamefont
  {Masciovecchio}, \citenamefont {Milne}, \citenamefont {Parmigiani},
  \citenamefont {Rugg}, \citenamefont {Spence}, \citenamefont {Thompson},
  \citenamefont {Ueda}, \citenamefont {Vinko}, \citenamefont {Wark},\ and\
  \citenamefont {Wurth}}]{Seddon2017}%
  \BibitemOpen
  \bibfield  {author} {\bibinfo {author} {\bibfnamefont {E.~A.}\ \bibnamefont
  {Seddon}}, \bibinfo {author} {\bibfnamefont {J.~A.}\ \bibnamefont {Clarke}},
  \bibinfo {author} {\bibfnamefont {D.~J.}\ \bibnamefont {Dunning}}, \bibinfo
  {author} {\bibfnamefont {C.}~\bibnamefont {Masciovecchio}}, \bibinfo {author}
  {\bibfnamefont {C.~J.}\ \bibnamefont {Milne}}, \bibinfo {author}
  {\bibfnamefont {F.}~\bibnamefont {Parmigiani}}, \bibinfo {author}
  {\bibfnamefont {D.}~\bibnamefont {Rugg}}, \bibinfo {author} {\bibfnamefont
  {J.~C.~H.}\ \bibnamefont {Spence}}, \bibinfo {author} {\bibfnamefont {N.~R.}\
  \bibnamefont {Thompson}}, \bibinfo {author} {\bibfnamefont {K.}~\bibnamefont
  {Ueda}}, \bibinfo {author} {\bibfnamefont {S.~M.}\ \bibnamefont {Vinko}},
  \bibinfo {author} {\bibfnamefont {J.~S.}\ \bibnamefont {Wark}},\ and\
  \bibinfo {author} {\bibfnamefont {W.}~\bibnamefont {Wurth}},\ }\bibfield
  {title} {\bibinfo {title} {Short-wavelength free-electron laser sources and
  science: a review},\ }\href {https://doi.org/10.1088/1361-6633/aa7cca}
  {\bibfield  {journal} {\bibinfo  {journal} {Rep. Prog. Phys.}\ }\textbf
  {\bibinfo {volume} {80}},\ \bibinfo {pages} {115901} (\bibinfo {year}
  {2017})}\BibitemShut {NoStop}%
\bibitem [{\citenamefont {Kling}\ \emph {et~al.}(2019)\citenamefont {Kling},
  \citenamefont {Giese}, \citenamefont {Carmesin}, \citenamefont {Sauerbrey},\
  and\ \citenamefont {Schleich}}]{PRA2019}%
  \BibitemOpen
  \bibfield  {author} {\bibinfo {author} {\bibfnamefont {P.}~\bibnamefont
  {Kling}}, \bibinfo {author} {\bibfnamefont {E.}~\bibnamefont {Giese}},
  \bibinfo {author} {\bibfnamefont {C.~M.}\ \bibnamefont {Carmesin}}, \bibinfo
  {author} {\bibfnamefont {R.}~\bibnamefont {Sauerbrey}},\ and\ \bibinfo
  {author} {\bibfnamefont {W.~P.}\ \bibnamefont {Schleich}},\ }\bibfield
  {title} {\bibinfo {title} {High-gain quantum free-electron laser: Emergence
  and exponential gain},\ }\href {https://doi.org/10.1103/PhysRevA.99.053823}
  {\bibfield  {journal} {\bibinfo  {journal} {Phys. Rev. A.}\ }\textbf
  {\bibinfo {volume} {99}},\ \bibinfo {pages} {053823} (\bibinfo {year}
  {2019})}\BibitemShut {NoStop}%
\bibitem [{\citenamefont {Bonifacio}\ and\ \citenamefont
  {Fares}(2016)}]{boni_epl}%
  \BibitemOpen
  \bibfield  {author} {\bibinfo {author} {\bibfnamefont {R.}~\bibnamefont
  {Bonifacio}}\ and\ \bibinfo {author} {\bibfnamefont {H.}~\bibnamefont
  {Fares}},\ }\bibfield  {title} {\bibinfo {title} {A fully quantum theory of
  high-gain free-electron laser},\ }\href
  {https://doi.org/10.1209/0295-5075/115/34004} {\bibfield  {journal} {\bibinfo
   {journal} {Europhys. Lett.}\ }\textbf {\bibinfo {volume} {115}},\ \bibinfo
  {pages} {34004} (\bibinfo {year} {2016})}\BibitemShut {NoStop}%
\bibitem [{\citenamefont {Bosco}\ \emph {et~al.}(1983)\citenamefont {Bosco},
  \citenamefont {Colson},\ and\ \citenamefont {Freedman}}]{bosco}%
  \BibitemOpen
  \bibfield  {author} {\bibinfo {author} {\bibfnamefont {P.}~\bibnamefont
  {Bosco}}, \bibinfo {author} {\bibfnamefont {W.}~\bibnamefont {Colson}},\ and\
  \bibinfo {author} {\bibfnamefont {R.}~\bibnamefont {Freedman}},\ }\bibfield
  {title} {\bibinfo {title} {Quantum/classical mode evolution in free electron
  laser oscillators},\ }\href {https://doi.org/10.1109/JQE.1983.1071871}
  {\bibfield  {journal} {\bibinfo  {journal} {IEEE J. Quantum Electron.}\
  }\textbf {\bibinfo {volume} {19}},\ \bibinfo {pages} {272} (\bibinfo {year}
  {1983})}\BibitemShut {NoStop}%
\bibitem [{\citenamefont {Bambini}\ and\ \citenamefont
  {Renieri}(1978)}]{bambi}%
  \BibitemOpen
  \bibfield  {author} {\bibinfo {author} {\bibfnamefont {A.}~\bibnamefont
  {Bambini}}\ and\ \bibinfo {author} {\bibfnamefont {A.}~\bibnamefont
  {Renieri}},\ }\bibfield  {title} {\bibinfo {title} {The free electron laser:
  A single-particle classical model},\ }\href
  {https://doi.org/10.1007/BF02762613} {\bibfield  {journal} {\bibinfo
  {journal} {Lett. Nuovo Cimento}\ }\textbf {\bibinfo {volume} {21}},\ \bibinfo
  {pages} {399 } (\bibinfo {year} {1978})}\BibitemShut {NoStop}%
\bibitem [{\citenamefont {Bambini}\ \emph {et~al.}(1979)\citenamefont
  {Bambini}, \citenamefont {Renieri},\ and\ \citenamefont {Stenholm}}]{brs}%
  \BibitemOpen
  \bibfield  {author} {\bibinfo {author} {\bibfnamefont {A.}~\bibnamefont
  {Bambini}}, \bibinfo {author} {\bibfnamefont {A.}~\bibnamefont {Renieri}},\
  and\ \bibinfo {author} {\bibfnamefont {S.}~\bibnamefont {Stenholm}},\
  }\bibfield  {title} {\bibinfo {title} {Classical theory of the free-electron
  laser in a moving frame},\ }\href {https://doi.org/10.1103/PhysRevA.19.2013}
  {\bibfield  {journal} {\bibinfo  {journal} {Phys. Rev. A}\ }\textbf {\bibinfo
  {volume} {19}},\ \bibinfo {pages} {2013 } (\bibinfo {year}
  {1979})}\BibitemShut {NoStop}%
\bibitem [{\citenamefont {Steiniger}\ \emph {et~al.}(2014)\citenamefont
  {Steiniger}, \citenamefont {Bussmann}, \citenamefont {Pausch}, \citenamefont
  {Cowan}, \citenamefont {Irman}, \citenamefont {Jochmann}, \citenamefont
  {Sauerbrey}, \citenamefont {Schramm},\ and\ \citenamefont
  {Debus}}]{steiniger}%
  \BibitemOpen
  \bibfield  {author} {\bibinfo {author} {\bibfnamefont {K.}~\bibnamefont
  {Steiniger}}, \bibinfo {author} {\bibfnamefont {M.}~\bibnamefont {Bussmann}},
  \bibinfo {author} {\bibfnamefont {R.}~\bibnamefont {Pausch}}, \bibinfo
  {author} {\bibfnamefont {T.}~\bibnamefont {Cowan}}, \bibinfo {author}
  {\bibfnamefont {A.}~\bibnamefont {Irman}}, \bibinfo {author} {\bibfnamefont
  {A.}~\bibnamefont {Jochmann}}, \bibinfo {author} {\bibfnamefont
  {R.}~\bibnamefont {Sauerbrey}}, \bibinfo {author} {\bibfnamefont
  {U.}~\bibnamefont {Schramm}},\ and\ \bibinfo {author} {\bibfnamefont
  {A.}~\bibnamefont {Debus}},\ }\bibfield  {title} {\bibinfo {title} {Optical
  free-electron lasers with traveling-wave {T}homson-scattering},\ }\href
  {https://doi.org/10.1088/0953-4075/47/23/234011} {\bibfield  {journal}
  {\bibinfo  {journal} {J. Phys. B: At. Mol. Opt. Phys.}\ }\textbf {\bibinfo
  {volume} {47}},\ \bibinfo {pages} {234011} (\bibinfo {year}
  {2014})}\BibitemShut {NoStop}%
\bibitem [{\citenamefont {Robb}\ and\ \citenamefont
  {Bonifacio}(2012)}]{robb2012}%
  \BibitemOpen
  \bibfield  {author} {\bibinfo {author} {\bibfnamefont {G.~R.~M.}\
  \bibnamefont {Robb}}\ and\ \bibinfo {author} {\bibfnamefont {R.}~\bibnamefont
  {Bonifacio}},\ }\bibfield  {title} {\bibinfo {title} {Coherent and
  spontaneous emission in the quantum free electron laser},\ }\href
  {https://doi.org/10.1063/1.4729337} {\bibfield  {journal} {\bibinfo
  {journal} {Phys. Plasmas}\ }\textbf {\bibinfo {volume} {19}},\ \bibinfo
  {pages} {073101} (\bibinfo {year} {2012})}\BibitemShut {NoStop}%
\bibitem [{\citenamefont {Louisell}\ \emph {et~al.}(1978)\citenamefont
  {Louisell}, \citenamefont {Lam},\ and\ \citenamefont {Copeland}}]{loui78}%
  \BibitemOpen
  \bibfield  {author} {\bibinfo {author} {\bibfnamefont {W.~H.}\ \bibnamefont
  {Louisell}}, \bibinfo {author} {\bibfnamefont {J.~F.}\ \bibnamefont {Lam}},\
  and\ \bibinfo {author} {\bibfnamefont {D.~A.}\ \bibnamefont {Copeland}},\
  }\bibfield  {title} {\bibinfo {title} {Effect of space charge on
  free-electron-laser gain},\ }\href {https://doi.org/10.1103/PhysRevA.18.655}
  {\bibfield  {journal} {\bibinfo  {journal} {Phys. Rev. A}\ }\textbf {\bibinfo
  {volume} {18}},\ \bibinfo {pages} {655 } (\bibinfo {year}
  {1978})}\BibitemShut {NoStop}%
\bibitem [{\citenamefont {Sprangle}\ and\ \citenamefont
  {Smith}(1980)}]{sprangle1}%
  \BibitemOpen
  \bibfield  {author} {\bibinfo {author} {\bibfnamefont {P.}~\bibnamefont
  {Sprangle}}\ and\ \bibinfo {author} {\bibfnamefont {R.~A.}\ \bibnamefont
  {Smith}},\ }\bibfield  {title} {\bibinfo {title} {Theory of free-electron
  lasers},\ }\href {https://doi.org/10.1103/PhysRevA.21.293} {\bibfield
  {journal} {\bibinfo  {journal} {Phys. Rev. A}\ }\textbf {\bibinfo {volume}
  {21}},\ \bibinfo {pages} {293 } (\bibinfo {year} {1980})}\BibitemShut
  {NoStop}%
\bibitem [{\citenamefont {Bogoliubov}\ and\ \citenamefont
  {Mitropolsky}(1961)}]{bogoliubov}%
  \BibitemOpen
  \bibfield  {author} {\bibinfo {author} {\bibfnamefont {N.~N.}\ \bibnamefont
  {Bogoliubov}}\ and\ \bibinfo {author} {\bibfnamefont {Y.~A.}\ \bibnamefont
  {Mitropolsky}},\ }\href@noop {} {\emph {\bibinfo {title} {Asymptotic Methods
  in the Theory of Non-Linear Oscillations}}}\ (\bibinfo  {publisher}
  {Hindustan Publishing Corporation, Delhi},\ \bibinfo {year}
  {1961})\BibitemShut {NoStop}%
\bibitem [{\citenamefont {Buishvili}\ \emph {et~al.}(1981)\citenamefont
  {Buishvili}, \citenamefont {Volzhan},\ and\ \citenamefont
  {Menabde}}]{higher}%
  \BibitemOpen
  \bibfield  {author} {\bibinfo {author} {\bibfnamefont {L.~L.}\ \bibnamefont
  {Buishvili}}, \bibinfo {author} {\bibfnamefont {E.~B.}\ \bibnamefont
  {Volzhan}},\ and\ \bibinfo {author} {\bibfnamefont {M.~G.}\ \bibnamefont
  {Menabde}},\ }\bibfield  {title} {\bibinfo {title} {Higher approximations in
  the theory of the average {H}amiltonian},\ }\href
  {https://doi.org/10.1007/BF01030852} {\bibfield  {journal} {\bibinfo
  {journal} {Teor. Mat. Fiz.}\ }\textbf {\bibinfo {volume} {46}},\ \bibinfo
  {pages} {166 } (\bibinfo {year} {1981})}\BibitemShut {NoStop}%
\bibitem [{\citenamefont {Schm\"{u}ser}\ \emph {et~al.}(2008)\citenamefont
  {Schm\"{u}ser}, \citenamefont {Dohlus},\ and\ \citenamefont
  {Rossbach}}]{schmueser}%
  \BibitemOpen
  \bibfield  {author} {\bibinfo {author} {\bibfnamefont {P.}~\bibnamefont
  {Schm\"{u}ser}}, \bibinfo {author} {\bibfnamefont {M.}~\bibnamefont
  {Dohlus}},\ and\ \bibinfo {author} {\bibfnamefont {J.}~\bibnamefont
  {Rossbach}},\ }\href@noop {} {\emph {\bibinfo {title} {Ultraviolet and Soft
  X-Ray Free-Electron Lasers}}}\ (\bibinfo  {publisher} {Springer,
  Heidelberg},\ \bibinfo {year} {2008})\BibitemShut {NoStop}%
\bibitem [{\citenamefont {Kling}(2018)}]{peter}%
  \BibitemOpen
  \bibfield  {author} {\bibinfo {author} {\bibfnamefont {P.}~\bibnamefont
  {Kling}},\ }\emph {\bibinfo {title} {Theory of the Free-Electron Laser: From
  Classical to Quantum}},\ \href {https://doi.org/10.18725/OPARU-5238} {Ph.D.
  thesis},\ \bibinfo  {school} {Universit{\"a}t Ulm} (\bibinfo {year}
  {2018})\BibitemShut {NoStop}%
\bibitem [{\citenamefont {Carmesin}\ \emph {et~al.}(2020)\citenamefont
  {Carmesin}, \citenamefont {Kling}, \citenamefont {Giese}, \citenamefont
  {Sauerbrey},\ and\ \citenamefont {Schleich}}]{carmesin20}%
  \BibitemOpen
  \bibfield  {author} {\bibinfo {author} {\bibfnamefont {C.~M.}\ \bibnamefont
  {Carmesin}}, \bibinfo {author} {\bibfnamefont {P.}~\bibnamefont {Kling}},
  \bibinfo {author} {\bibfnamefont {E.}~\bibnamefont {Giese}}, \bibinfo
  {author} {\bibfnamefont {R.}~\bibnamefont {Sauerbrey}},\ and\ \bibinfo
  {author} {\bibfnamefont {W.~P.}\ \bibnamefont {Schleich}},\ }\bibfield
  {title} {\bibinfo {title} {Quantum and classical phase-space dynamics of a
  free-electron laser},\ }\href
  {https://doi.org/10.1103/PhysRevResearch.2.023027} {\bibfield  {journal}
  {\bibinfo  {journal} {Phys. Rev. Research}\ }\textbf {\bibinfo {volume}
  {2}},\ \bibinfo {pages} {023027} (\bibinfo {year} {2020})}\BibitemShut
  {NoStop}%
\bibitem [{\citenamefont {Kunze}\ \emph {et~al.}(1996)\citenamefont {Kunze},
  \citenamefont {D\"urr},\ and\ \citenamefont {Rempe}}]{Kunze1996}%
  \BibitemOpen
  \bibfield  {author} {\bibinfo {author} {\bibfnamefont {S.}~\bibnamefont
  {Kunze}}, \bibinfo {author} {\bibfnamefont {S.}~\bibnamefont {D\"urr}},\ and\
  \bibinfo {author} {\bibfnamefont {G.}~\bibnamefont {Rempe}},\ }\bibfield
  {title} {\bibinfo {title} {Bragg scattering of slow atoms from a standing
  light wave},\ }\href {https://doi.org/10.1209/epl/i1996-00462-x} {\bibfield
  {journal} {\bibinfo  {journal} {Europhys. Lett.}\ }\textbf {\bibinfo {volume}
  {34}},\ \bibinfo {pages} {343} (\bibinfo {year} {1996})}\BibitemShut
  {NoStop}%
\bibitem [{\citenamefont {Ahlers}\ \emph {et~al.}(2016)\citenamefont {Ahlers},
  \citenamefont {M\"untinga}, \citenamefont {Wenzlawski}, \citenamefont
  {Krutzik}, \citenamefont {Tackmann}, \citenamefont {Abend}, \citenamefont
  {Gaaloul}, \citenamefont {Giese}, \citenamefont {Roura}, \citenamefont
  {Kuhl}, \citenamefont {L\"ammerzahl}, \citenamefont {Peters}, \citenamefont
  {Windpassinger}, \citenamefont {Sengstock}, \citenamefont {Schleich},
  \citenamefont {Ertmer},\ and\ \citenamefont {Rasel}}]{Ahlers2016}%
  \BibitemOpen
  \bibfield  {author} {\bibinfo {author} {\bibfnamefont {H.}~\bibnamefont
  {Ahlers}}, \bibinfo {author} {\bibfnamefont {H.}~\bibnamefont {M\"untinga}},
  \bibinfo {author} {\bibfnamefont {A.}~\bibnamefont {Wenzlawski}}, \bibinfo
  {author} {\bibfnamefont {M.}~\bibnamefont {Krutzik}}, \bibinfo {author}
  {\bibfnamefont {G.}~\bibnamefont {Tackmann}}, \bibinfo {author}
  {\bibfnamefont {S.}~\bibnamefont {Abend}}, \bibinfo {author} {\bibfnamefont
  {N.}~\bibnamefont {Gaaloul}}, \bibinfo {author} {\bibfnamefont
  {E.}~\bibnamefont {Giese}}, \bibinfo {author} {\bibfnamefont
  {A.}~\bibnamefont {Roura}}, \bibinfo {author} {\bibfnamefont
  {R.}~\bibnamefont {Kuhl}}, \bibinfo {author} {\bibfnamefont {C.}~\bibnamefont
  {L\"ammerzahl}}, \bibinfo {author} {\bibfnamefont {A.}~\bibnamefont
  {Peters}}, \bibinfo {author} {\bibfnamefont {P.}~\bibnamefont
  {Windpassinger}}, \bibinfo {author} {\bibfnamefont {K.}~\bibnamefont
  {Sengstock}}, \bibinfo {author} {\bibfnamefont {W.~P.}\ \bibnamefont
  {Schleich}}, \bibinfo {author} {\bibfnamefont {W.}~\bibnamefont {Ertmer}},\
  and\ \bibinfo {author} {\bibfnamefont {E.~M.}\ \bibnamefont {Rasel}},\
  }\bibfield  {title} {\bibinfo {title} {Double {B}ragg interferometry},\
  }\href {https://doi.org/10.1103/PhysRevLett.116.173601} {\bibfield  {journal}
  {\bibinfo  {journal} {Phys. Rev. Lett.}\ }\textbf {\bibinfo {volume} {116}},\
  \bibinfo {pages} {173601} (\bibinfo {year} {2016})}\BibitemShut {NoStop}%
\bibitem [{\citenamefont {Bonifacio}\ \emph {et~al.}(1986)\citenamefont
  {Bonifacio}, \citenamefont {Casagrande},\ and\ \citenamefont
  {De~Salvo~Souza}}]{boni_coll}%
  \BibitemOpen
  \bibfield  {author} {\bibinfo {author} {\bibfnamefont {R.}~\bibnamefont
  {Bonifacio}}, \bibinfo {author} {\bibfnamefont {F.}~\bibnamefont
  {Casagrande}},\ and\ \bibinfo {author} {\bibfnamefont {L.}~\bibnamefont
  {De~Salvo~Souza}},\ }\bibfield  {title} {\bibinfo {title} {Collective
  variable description of a free-electron laser},\ }\href
  {https://doi.org/10.1103/PhysRevA.33.2836} {\bibfield  {journal} {\bibinfo
  {journal} {Phys. Rev. A}\ }\textbf {\bibinfo {volume} {33}},\ \bibinfo
  {pages} {2836 } (\bibinfo {year} {1986})}\BibitemShut {NoStop}%
\bibitem [{\citenamefont {Dicke}(1954)}]{dicke}%
  \BibitemOpen
  \bibfield  {author} {\bibinfo {author} {\bibfnamefont {R.~H.}\ \bibnamefont
  {Dicke}},\ }\bibfield  {title} {\bibinfo {title} {Coherence in spontaneous
  radiation processes},\ }\href {https://doi.org/10.1103/PhysRev.93.99}
  {\bibfield  {journal} {\bibinfo  {journal} {Phys. Rev.}\ }\textbf {\bibinfo
  {volume} {93}},\ \bibinfo {pages} {99 } (\bibinfo {year} {1954})}\BibitemShut
  {NoStop}%
\bibitem [{\citenamefont {Byrd}\ and\ \citenamefont {Friedman}(1971)}]{byrd}%
  \BibitemOpen
  \bibfield  {author} {\bibinfo {author} {\bibfnamefont {P.~F.}\ \bibnamefont
  {Byrd}}\ and\ \bibinfo {author} {\bibfnamefont {M.~D.}\ \bibnamefont
  {Friedman}},\ }\href@noop {} {\emph {\bibinfo {title} {Handbook of Elliptic
  Integrals for Engineers and Scientists}}}\ (\bibinfo  {publisher} {Springer,
  Berlin},\ \bibinfo {year} {1971})\BibitemShut {NoStop}%
\bibitem [{\citenamefont {Compagno}\ \emph {et~al.}(1986)\citenamefont
  {Compagno}, \citenamefont {Peng},\ and\ \citenamefont
  {Persico}}]{compagno86}%
  \BibitemOpen
  \bibfield  {author} {\bibinfo {author} {\bibfnamefont {G.}~\bibnamefont
  {Compagno}}, \bibinfo {author} {\bibfnamefont {J.~S.}\ \bibnamefont {Peng}},\
  and\ \bibinfo {author} {\bibfnamefont {F.}~\bibnamefont {Persico}},\
  }\bibfield  {title} {\bibinfo {title} {Squeezing in a two-photon {D}icke
  hamiltonian},\ }\href
  {https://doi.org/https://doi.org/10.1016/0030-4018(86)90221-X} {\bibfield
  {journal} {\bibinfo  {journal} {Opt. Commun.}\ }\textbf {\bibinfo {volume}
  {57}},\ \bibinfo {pages} {415} (\bibinfo {year} {1986})}\BibitemShut
  {NoStop}%
\bibitem [{\citenamefont {Gerry}\ and\ \citenamefont {Togeas}(1989)}]{gerry89}%
  \BibitemOpen
  \bibfield  {author} {\bibinfo {author} {\bibfnamefont {C.~C.}\ \bibnamefont
  {Gerry}}\ and\ \bibinfo {author} {\bibfnamefont {J.~B.}\ \bibnamefont
  {Togeas}},\ }\bibfield  {title} {\bibinfo {title} {Squeezing and photon
  antibunching from a two-photon dicke model},\ }\href
  {https://doi.org/https://doi.org/10.1016/0030-4018(89)90112-0} {\bibfield
  {journal} {\bibinfo  {journal} {Opt. Commun.}\ }\textbf {\bibinfo {volume}
  {69}},\ \bibinfo {pages} {263} (\bibinfo {year} {1989})}\BibitemShut
  {NoStop}%
\bibitem [{\citenamefont {Kumar}\ and\ \citenamefont {Mehta}(1980)}]{kumar}%
  \BibitemOpen
  \bibfield  {author} {\bibinfo {author} {\bibfnamefont {S.}~\bibnamefont
  {Kumar}}\ and\ \bibinfo {author} {\bibfnamefont {C.~L.}\ \bibnamefont
  {Mehta}},\ }\bibfield  {title} {\bibinfo {title} {Theory of the interaction
  of a single-mode resonant radiation field with \textit{{N}} two-level
  atoms},\ }\href {https://doi.org/10.1103/PhysRevA.21.1573} {\bibfield
  {journal} {\bibinfo  {journal} {Phys. Rev. A}\ }\textbf {\bibinfo {volume}
  {21}},\ \bibinfo {pages} {1573 } (\bibinfo {year} {1980})}\BibitemShut
  {NoStop}%
\bibitem [{Note1()}]{Note1}%
  \BibitemOpen
  \bibinfo {note} {For a small interaction length $L$ we find the asymptotic
  behavior $n\cong n_0 \left [1+n_0\left (\alpha _N L/L_g\right )^2/(2N)\right
  ]$. Hence, a linear analysis (often used in FEL theory) is not sufficient to
  obtain this non-linear short-time behavior.}\BibitemShut {Stop}%
\bibitem [{\citenamefont {Walls}\ and\ \citenamefont
  {Barakat}(1970)}]{walls70}%
  \BibitemOpen
  \bibfield  {author} {\bibinfo {author} {\bibfnamefont {D.~F.}\ \bibnamefont
  {Walls}}\ and\ \bibinfo {author} {\bibfnamefont {R.}~\bibnamefont
  {Barakat}},\ }\bibfield  {title} {\bibinfo {title} {Quantum-mechanical
  amplification and frequency conversion with a trilinear {H}amiltonian},\
  }\href {https://doi.org/10.1103/PhysRevA.1.446} {\bibfield  {journal}
  {\bibinfo  {journal} {Phys. Rev. A}\ }\textbf {\bibinfo {volume} {1}},\
  \bibinfo {pages} {446 } (\bibinfo {year} {1970})}\BibitemShut {NoStop}%
\bibitem [{Note2()}]{Note2}%
  \BibitemOpen
  \bibinfo {note} {Note that the results in the high-gain regime, Eqs.~\protect
  \textup {\hbox {\mathsurround \z@ \protect \normalfont (\ignorespaces \ref
  {eq:nkl_erste}\unskip \@@italiccorr )}} and~\protect \textup {\hbox
  {\mathsurround \z@ \protect \normalfont (\ignorespaces \ref
  {eq:qhigh_nkl_zweite}\unskip \@@italiccorr )}}, reduce to their respective
  low-gain counterparts in Eq.~\protect \textup {\hbox {\mathsurround \z@
  \protect \normalfont (\ignorespaces \ref {eq:dn_low}\unskip \@@italiccorr )}}
  in the asymptotic limit $\delta n \sim N \ll n_0$.}\BibitemShut {Stop}%
\bibitem [{\citenamefont {Bonifacio}(2005)}]{boni05}%
  \BibitemOpen
  \bibfield  {author} {\bibinfo {author} {\bibfnamefont {R.}~\bibnamefont
  {Bonifacio}},\ }\bibfield  {title} {\bibinfo {title} {Quantum {SASE} {FEL}
  with laser wiggler},\ }\href {https://doi.org/10.1016/j.nima.2005.04.003}
  {\bibfield  {journal} {\bibinfo  {journal} {Nucl. Instrum. \& Methods A}\
  }\textbf {\bibinfo {volume} {546}},\ \bibinfo {pages} {634 } (\bibinfo {year}
  {2005})}\BibitemShut {NoStop}%
\bibitem [{\citenamefont {Nayfeh}(1973)}]{nayfeh}%
  \BibitemOpen
  \bibfield  {author} {\bibinfo {author} {\bibfnamefont {A.~H.}\ \bibnamefont
  {Nayfeh}},\ }\href@noop {} {\emph {\bibinfo {title} {Perturbation Methods}}}\
  (\bibinfo  {publisher} {Wiley, New York},\ \bibinfo {year}
  {1973})\BibitemShut {NoStop}%
\bibitem [{\citenamefont {Schwinger}(1952)}]{schwinger}%
  \BibitemOpen
  \bibfield  {author} {\bibinfo {author} {\bibfnamefont {J.}~\bibnamefont
  {Schwinger}},\ }\href@noop {} {\emph {\bibinfo {title} {{O}n {A}ngular
  {M}omentum}}},\ \bibinfo {type} {Tech. Rep.}\ (\bibinfo  {institution}
  {United States Atomic Energy Comission},\ \bibinfo {year} {1952})\BibitemShut
  {NoStop}%
\bibitem [{\citenamefont {Cohen-Tannoudji}\ \emph {et~al.}(1977)\citenamefont
  {Cohen-Tannoudji}, \citenamefont {Diu},\ and\ \citenamefont
  {Lalo{\"e}}}]{ct}%
  \BibitemOpen
  \bibfield  {author} {\bibinfo {author} {\bibfnamefont {C.}~\bibnamefont
  {Cohen-Tannoudji}}, \bibinfo {author} {\bibfnamefont {B.}~\bibnamefont
  {Diu}},\ and\ \bibinfo {author} {\bibfnamefont {F.}~\bibnamefont
  {Lalo{\"e}}},\ }\href@noop {} {\emph {\bibinfo {title} {Quantum Mechanics,
  Volume One}}}\ (\bibinfo  {publisher} {Wiley, Singapore},\ \bibinfo {year}
  {1977})\BibitemShut {NoStop}%
\end{thebibliography}

\end{document}